\documentclass[11pt,oneside]{article}
\usepackage{a4wide}
\usepackage{mathrsfs}
\usepackage{latexsym,bm}
\usepackage{graphicx}
\usepackage{indentfirst}
\usepackage{slashed}
\usepackage{amsmath}
\usepackage{amssymb}
\usepackage{color}
\usepackage{hyperref}
\usepackage{epsfig}
\usepackage[titletoc]{appendix}
\usepackage{multirow}%
\usepackage{rotating}
\usepackage{cite}
\usepackage{ulem} 
\usepackage[top=2.9cm,bottom=2.5cm,left=2.8cm,right=3cm]{geometry}
\usepackage{tabularx}

\newcommand{\be}{\begin{equation}} \newcommand{\ee}{\end{equation}}
\newcommand{\ba}{\left(\begin{array}{c}}
\newcommand{\ea}{\end{array}\right)}
\newcommand{\bea}{\begin{eqnarray}} \newcommand{\eea}{\end{eqnarray}}

\newcommand{\al}{&\!\!\!\!}

\newcommand{\ds}{D_{s0}^*(2317)}
\newcommand{\dd}{D_{0}^*(2400)}

\newcommand{\bma}{\left(\begin{matrix}}
\newcommand{\ema}{\end{matrix}\right)}
\newcommand{\bqa}{\begin{eqnarray}}
\newcommand{\eqa}{\end{eqnarray}}
\newcommand{\bqaa}{\begin{eqnarray*}}
\newcommand{\eqaa}{\end{eqnarray*}}

\newcommand{\bra}{\langle}
\newcommand{\ket}{\rangle}
\newcommand{\nn}{\nonumber}
\newcommand{\mD}{\mathcal{D}}
\newcommand{\mP}{\mathcal{P}}
\newcommand{\mV}{\mathcal{V}}

\begin{document}
\thispagestyle{empty}
\title{
\Large \bf Towards a precise determination of the scattering amplitudes of the charmed and light-flavor pseudoscalar mesons }
\author{\small Zhi-Hui Guo$^{a}$, \,  Liuming Liu$^{b}$,  \,
Ulf-G. Mei{\ss}ner$^{c,d,e}$, \, J.~A.~Oller$^{f}$,\, A.~Rusetsky$^{c}$   \\[0.5em]
{ \small\it ${}^a$  Department of Physics and Hebei Advanced Thin Films Laboratory, } \\
{\small\it Hebei Normal University,  Shijiazhuang 050024, China}\\[0.3em]
{ \small\it ${}^b$ Institute of Modern Physics, Chinese Academy of Sciences, Lanzhou 730000, China}\\[0.3em]
{\small\it  $^c$Helmholtz-Institut f\"ur Strahlen- und Kernphysik and
Bethe Center for Theoretical Physics,}\\
{\small\it Universit\"at Bonn, D--53115
Bonn, Germany}\\[0.3em]
{\small\it  $^d$Institute for Advanced Simulation, Institut f{\"u}r
Kernphysik and J\"ulich Center for Hadron Physics,}\\
{\small\it Forschungszentrum  J{\"u}lich, D-52425 J{\"u}lich, Germany}\\
{\small {\it $^e$ Ivane Javakhishvili Tbilisi State University, 0186 Tbilisi, Georgia}}\\
{\small {\it $^f$Departamento de F\'{\i}sica. Universidad de Murcia. E-30071 Murcia. Spain}}
}
\date{}
\maketitle
\begin{abstract}
We study the scattering of the light-flavor pseudoscalar mesons ($\pi, K, \eta$) off the ground-state charmed mesons ($D,D_s$) within chiral effective field theory. The recent lattice simulation results on various scattering lengths and the finite-volume spectra both in the moving and center-of-mass frames, most of which are obtained at unphysical meson masses, are used to constrain the free parameters in our theory. Explicit formulas to include the $S$- and $P$-wave mixing to determine the finite-volume energy levels are provided. After a successful reproduction of the lattice data, we perform a 
chiral extrapolation to predict the quantities with physical meson masses, including phase shifts, inelasticities, resonance pole positions and the corresponding residues from the scattering of the light pseudoscalar and charmed mesons.

\end{abstract}




\section{Introduction}

The spectroscopy of the open charmed mesons is an active and interesting research topic in hadron physics. The  discovery
of the scalar charm-strange meson $\ds$ ~\cite{Aubert:2003fg,Besson:2003cp,Krokovny:2003zq} challenges the quark
model description~\cite{Godfrey:1985xj}, which predicts a mass around 160~MeV heavier than the experimental
value. Another puzzle is that the mass of $\ds$ is almost the same as the mass of its non-strange partner $\dd$. 
The scattering process of the ground-state charmed mesons ($D,D_s$) and the light pseudoscalar mesons ($\pi, K, \eta$)
offers an excellent environment to explore the properties of the scalar charmed resonances $\ds$, $\dd$
and possible resonances with other quantum numbers as well.

Chiral effective field theory provides a useful theoretical framework to perform such studies. Many works along
this research line have been done by several different groups in the last  decade~\cite{Kolomeitsev:2003ac,Hofmann:2003je,Guo:2008gp,Guo:2006fu,Guo:2009ct,Cleven:2010aw,MartinezTorres:2014vna,MartinezTorres:2011pr,Yao:2015qia,Guo:2015dha,Albaladejo:2016lbb,Du:2017ttu,Guo:2018kno,Albaladejo:2018mhb,Altenbuchinger:2013gaa,Altenbuchinger:2013vwa,Geng:2010vw,Wang:2012bu,Liu:2009uz,Liu:2012zya}. 
In order to constrain the unknown parameters, one usually needs scattering information as input. However, experimental
observables from the scattering of the light pseudoscalar and ground-state charmed mesons, such as phase shifts and
inelasticities, are still not available nowadays. 

Fortunately, lattice QCD provides an alternative way to obtain such kinds of data~\cite{Liu:2012zya,Mohler:2013rwa,Lang:2014yfa,Bali:2017pdv,Moir:2016srx, Mohler:2012na}.  In Ref.~\cite{Liu:2012zya}, the scattering lengths of five scattering channels: 
isospin-3/2 $D\pi$, $D_s \pi$, $D_s K$, isospin-0 $D\bar{K}$ and isospin-1 $D\bar{K}$, are calculated at four
different values of unphysical pion (quark) masses. The $DK$ scattering amplitude is obtained indirectly from
unitarized chiral perturbation theory (ChPT) with the relevant low-energy constants (LECs) determined from the
aforementioned five channels. The direct lattice calculation of $DK$ scattering is performed in
Refs.~\cite{Mohler:2013rwa,Lang:2014yfa,Bali:2017pdv} and the  $I=1/2$ $D\pi$ scattering length is calculated in
lattice QCD in Ref.~\cite{Mohler:2012na}. In these works, the effects of the coupled channels are ignored.
Recently, a sophisticated lattice calculation of the coupled-channel scattering of $D\pi, D\eta$ and $D_s \bar{K}$
was presented in Ref.~\cite{Moir:2016srx}, in which a large amount of energy levels in the finite volume are obtained by
using many interpolating operators and various moving frames. These lattice data have been extensively used in
ChPT studies to constrain the chiral amplitudes~\cite{Yao:2015qia,Guo:2015dha,Du:2017ttu,Altenbuchinger:2013gaa,Altenbuchinger:2013vwa,Wang:2012bu,Liu:2012zya, Albaladejo:2016lbb}. However, all of these studies used only a small part of the 
available lattice data up to now. A more complete data set is expected to be able to determine the chiral amplitudes
more precisely. In this work, we perform an extensive study of the light pseudoscalar mesons scattering off the
ground-state charmed mesons in unitarized ChPT. All of the 2+1 flavor lattice results, including the finite-volume
energy levels and the scattering lengths obtained in Ref.~\cite{Liu:2012zya,Mohler:2013rwa,Lang:2014yfa, Moir:2016srx},
are used to determine the parameters in the unitarized ChPT. Note that the 2-flavor lattice data in
Refs.~\cite{Mohler:2013rwa,Lang:2014yfa,Bali:2017pdv, Mohler:2012na} are not used in our analysis. We follow the
theoretical framework in Refs.~\cite{Doring:2011vk,Doring:2012eu,Gockeler:2012yj} to analyze the lattice finite-volume
spectra. The essential difference between this approach and the $K$-matrix assisted L\"uscher method~\cite{Luscher:1990ux}
used in Ref.~\cite{Moir:2016srx} is that Ref.~\cite{Moir:2016srx} relies on a given algebraic parameterization of
the $K$-matrix, whereas in this paper the scattering amplitude (and hence the multichannel $K$-matrix)
is obtained through the solution of dynamical equations with the kernel calculated in ChPT. Therefore, not only can we
extract the scattering parameters and the resonance properties at unphysical meson masses, but we can also predict
these quantities at physical meson masses by performing a chiral extrapolation.

This article is organized as follows. The relevant chiral Lagrangians, the perturbative scattering amplitudes and their
unitarization are discussed in Sec.~\ref{sect.chptamp}. The finite-volume effects in the chiral effective field theory
are elaborated on in Sec.~\ref{sect.uniandfv}. The fits to the finite-volume spectra and the scattering lengths
are presented in Sec.~\ref{sect.fit}. The scattering phase shifts, inelasticities, resonance pole positions and
the residues are discussed in detail in Sec.~\ref{sect.pheno}. A short summary and conclusions are given in
Sec.~\ref{sect.conclusion}.

\section{Chiral amplitudes and  unitarization}\label{sect.chptamp} 

We take into account the chiral Lagrangians involving the light pseudoscalar and the ground-state charmed mesons
up to next-to-leading order (NLO). Detailed discussions on chiral Lagrangians up to next-to-next-to-leading order
can be found in Refs.~\cite{Yao:2015qia,Du:2017ttu}. In the $SU(3)$ chiral Lagrangian the octet state $\eta_8$ is
identified as the physical $\eta$ meson~\cite{Gasser:1984gg}. The $U(3)$ chiral theory allows one to simultaneously
include the physical $\eta$ and $\eta'$ mesons, by explicitly incorporating the singlet  $\eta_0$~\cite{Kaiser:2000gs}.
The generalization of the $U(3)$ chiral study in the scattering of the charmed and the light pseudoscalar mesons
is carried out in Ref.~\cite{Guo:2015dha}. It is found that the massive $\eta'$ meson plays a minor role in the energy
region considered, therefore we work in the conventional $SU(3)$ chiral Lagrangian in this work.

We briefly introduce the relevant $SU(3)$ chiral Lagrangians to set up our notations. The ground-state charmed-meson
triplet $\mP=(D^0,D^+,D^+_{s})$ is incorporated in the chiral Lagrangians as a matter field. The light pseudoscalar
mesons $\pi, K$ and $\eta$ are treated as pseudo-Nambu-Goldstone bosons (pNGBs). The leading-order (LO) chiral
Lagrangian describing the interactions between the pNGBs and the charmed mesons reads 
\begin{eqnarray}\label{lolag}
\mathcal{L}^{(1)}_{\mP\phi}=\mathcal{D}_\mu \mP \mathcal{D}^\mu \mP^\dagger-\overline{M}_D^2 \mP \mP^\dagger\ ,
\end{eqnarray}
where $\overline{M}_D$ denotes the mass of the charmed-meson triplet in the chiral limit. 
The covariant derivative $\mD_\mu$ is given by 
\begin{eqnarray}
\mathcal{D}_\mu \mP= \mP(\overset{\leftarrow}{\partial_\mu}+\Gamma_\mu^\dagger)\ ,
\qquad \mathcal{D}_\mu \mP^\dagger=(\partial_\mu+\Gamma_\mu) \mP^\dagger\ ,
\end{eqnarray}
where 
\begin{eqnarray}\label{defbb}
\Gamma_\mu  &=& \frac{1}{2}\bigg( u^\dagger \partial_\mu u
+ u \partial_\mu  u^\dagger \bigg)\,, \nn\\
u^2 &=& e^{i\frac{ \sqrt2\Phi}{ F}} \,, \nn \\
\Phi &=&   \bma
      \frac{1}{\sqrt{2}}\pi^0+\frac{1}{\sqrt{6}}\eta_8  & \pi^+ &K^+ \\
      \pi^- & -\frac{1}{\sqrt{2}}\pi^0+\frac{1}{\sqrt{6}}\eta_8 &K^0\\
      K^-& \overline{K}^0&\frac{-2}{\sqrt{6}}\eta_8\\
   \ema\ .
\end{eqnarray}
Here, $F$ denotes the weak decay constant of the pNGBs in the chiral limit, with the normalization $F_\pi=92.1$~MeV. 
The NLO Lagrangian, with six additional low energy constants $h_{i=0,\ldots,5}$, takes the form~\cite{Guo:2008gp,Guo:2009ct}
\begin{eqnarray}\label{nlolag}
\mathcal{L}^{(2)}_{\mP\phi} &=& \mP \bigg(-h_0\langle\chi_+\rangle-h_1{\chi}_+
+ h_2\langle u_\mu u^\mu\rangle-h_3u_\mu u^\mu \bigg) {\mP}^\dag  \nn \\&&
+ \mathcal{D}_\mu \mP \bigg( {h_4}\langle u_\mu
u^\nu\rangle-{h_5}\{u^\mu,u^\nu\} \bigg)\mathcal{D}_\nu {\mP}^\dag\,, 
\end{eqnarray}
with  
\begin{eqnarray}\label{defbb2}
\chi_+ \,=\, u^\dagger  \chi u^\dagger  +  u \chi^\dagger  u \,, \qquad
u_\mu \, =\,
i ( u^\dagger \partial_\mu   u\, -\, u \partial_\mu   u^\dagger )\,, \qquad \chi=2 B (s + i p)\,,
\end{eqnarray}
where $s$ and $p$ denote the scalar and pseudoscalar external sources, respectively. By taking $(s + i p)$
= diag($\hat{m},\hat{m},m_s$), with $\hat{m}$ the average of up- and down-quark mass and $m_s$ the strange quark mass,
one can introduce the light-quark masses in the chiral Lagrangian. We do not consider any isospin violation effect
in this work. At leading order, the quantity $B$ in Eq.~\eqref{defbb2} is related to the light-quark condensate
through $\bra 0| \bar{q}^i q^j|0\ket = -F^2 B\delta^{ij}$.  The LO squared masses of the pNGBs are then given by 
\begin{equation}\label{mpimq}
 m_\pi^2 = 2 B \hat{m}\,, \qquad   m_K^2 = B (\hat{m} + m_s )\,, \qquad  m_\eta^2 = \frac{4m_K^2-m_\pi^2}{3}\,.
\end{equation}

With different combinations of the strangeness ($S$) and isospin ($I$), the scattering amplitudes of the ground-state
charmed mesons and the pNGBs are classified into seven different cases. See the first and second columns of Table~\ref{tab:ci}
for the specific channels involved in each case. For the process $D_1(p_1) + \phi_1(p_2) \to D_2 (p_3) + \phi_2(p_4)$
with definite strangeness and isospin, the general scattering amplitude takes the form 
\begin{equation}\label{eq.v}
V^{(S,I)}_{D_1\phi_1\to D_2\phi_2}(s,t,u) = \frac{1}{F_\pi^2} \bigg[\frac{C_{\rm LO}}{4}(s-u) - 4 C_0 h_0 +
2 C_1 h_1 - 2C_{24} H_{24}(s,t,u) + 2C_{35} H_{35}(s,t,u) 
\bigg]\,,  
\end{equation}
where $s=(p_1+p_2)^2=(p_3+p_4)^2,t=(p_1-p_3)^2=(p_4-p_2)^2,u=(p_1-p_4)^2=(p_3-p_2)^2$ correspond to the standard Mandelstam variables, and the functions $H_{24}(s,t,u)$ and $H_{35}(s,t,u)$ are given by
\begin{eqnarray}
H_{24}(s,t,u) &=& 2 h_2\, (p_2\cdot p_4) + h_4 \,[ (p_1\cdot p_2) (p_3\cdot p_4) +
(p_1\cdot p_4)( p_2\cdot p_3)]\,, \\
H_{35}(s,t,u) &=& h_3 \, (p_2\cdot p_4) + h_5 \,
[(p_1\cdot p_2)( p_3\cdot p_4) + (p_1\cdot p_4)( p_2\cdot p_3)]\,.
\end{eqnarray}
The coefficients $C_i$ in Eq.~\eqref{eq.v} have been given in many  works~\cite{Guo:2009ct,Geng:2010vw,Wang:2012bu} and
we show their expressions in Table~\ref{tab:ci} for the sake of completeness. The results from the generalization to the
$U(3)$ case with explicit $\eta'$ meson have been given in Ref.~\cite{Guo:2015dha}.

\begin{table}[tbph]
\centering
\begin{tabular}{|l c | c c c c c | }
\hline\hline
 $(S,I)$ & Channels & $C_{\rm LO}$ & $C_0$ & $C_1$ & $C_{24}$ & $C_{35}$ 
 \\
\hline
$(-1,0)$      & $D\bar{K}\to D\bar K$   & $-1$  & $m_K^2$ & $m_K^2$  & 1 & $-1$  
\\
\hline
$(-1,1)$      & $D\bar{K}\to D\bar K$   & 1     & $m_K^2$ & $-m_K^2$ & 1 & 1 
\\
\hline
$(2,\frac12)$ & $D_sK\to D_sK$          & 1     & $m_K^2$ & $-m_K^2$ & 1 & 1  
\\
\hline
$(0,\frac32)$ & $D\pi\to D\pi$          & 1 & $m_\pi^2$ & $-m_\pi^2$ & 1 & 1  
\\
\hline
$(1,1)$       & $D_s\pi\to D_s\pi$      & 0   & $m_\pi^2$ & 0        & 1 & 0   
\\
              & $D K\to D K$            & 0     & $m_K^2$ & 0        & 1 & 0 
\\
              & $D K\to D_s\pi$   & 1 & 0 & $-(m_K^2+m_\pi^2)/2$ & 0    & 1 
\\
\hline
$(1,0)$       & $D K\to D K$            & $-2$ & $m_K^2$ & $-2m_K^2$ & 1 & 2 
\\
             & $D K\to D_s\eta$ & $-\sqrt{3}$ & 0 &
        $ \frac{-5m_K^2+3m_\pi^2}{2\sqrt{3}} $ & 0 
        & $\frac{1}{\sqrt{3}} $  
\\
              & $D_s\eta\to D_s\eta$ & 0 & $ \frac{4m_K^2-m_\pi^2}{3}$ 
              &  $ \frac{4(m_\pi^2-2m_K^2)}{3} $
             & 1 & $ \frac{4}{3}$  
             
\\ 
\hline 
$(0,\frac12)$ & $D\pi\to D\pi$       & $-2$    & $m_\pi^2$ & $-m_\pi^2$ & 1 & 1 
\\
             & $D\eta\to D\eta$     & 0 & $  \frac{4m_K^2-m_\pi^2}{3} $ & $ \frac{-m_\pi^2}{3} $& 1
             & $ \frac{1}{3} $  
\\
             & $D_s\bar K\to D_s\bar K$& $-1$& $m_K^2$& $-m_K^2$& 1 & 1  
\\ 
             & $D\eta\to D\pi$     & 0 & 0 & $-m_\pi^2$ & 0 & $1$  
\\
             & $D_s\bar K\to D\pi$ & $-\frac{\sqrt{6}}{2}$ & 0 &
             $\frac{-{\sqrt{6}}(m_K^2+m_\pi^2)}{4}$ & 0 & $\frac{\sqrt{6}}{2}$   
\\
             & $D_s\bar K\to D\eta$& $-\frac{\sqrt{6}}{2}$ & 0 &
             $  \frac{5m_K^2-3m_\pi^2}{2\sqrt{6}} $ & 0 & $ \frac{-1}{\sqrt{6}}$ 
\\
\hline\hline
\end{tabular}
\caption{\label{tab:ci} The coefficients $C_i$ in the amplitudes $V^{(S,I)}_{D_1\phi_1\to D_2\phi_2}(s,t,u)$ of Eq.~\eqref{eq.v}.
The quantum numbers of different channels are classified by strangeness ($S$) and isospin ($I$), as shown in the first column. } 
\end{table}

In the present work we mainly focus on the $S$-wave scattering of the pNGBs and the charmed mesons. 
In order to obtain the partial-wave amplitudes, we need to perform the partial-wave projection of the full
amplitudes in Eq.~\eqref{eq.v} with angular momentum $J$. The explicit formula reads
\begin{eqnarray}\label{eq.pwv}
\mathcal{V}_{J,\,D_1\phi_1\to D_2\phi_2}^{(S,I)}(s)
= \frac{1}{2}\int_{-1}^{+1}{\rm 
d}\cos\varphi\,P_J(\cos\varphi)\, V^{(S,I)}_{D_1\phi_1\to
D_2\phi_2}(s,t(s,\cos\varphi))\,,
\end{eqnarray}
where $\varphi$ is the scattering angle of the incoming and outgoing states  
in the center-of-mass (CM) frame, and the Mandelstam variable $t$ is related to $\varphi$ through
\begin{eqnarray}
t(s,\cos\varphi)\al=\al m_{D_1}^2+m_{D_2}^2-
\frac{1}{2s}\left(s+m_{D_1}^2-m_{\phi_1}^2\right)
\left(s+m_{D_2}^2-m_{\phi_2}^2\right) \nonumber\\
\al\al
-\frac{\cos\varphi}{2s}\sqrt{\lambda(s,m_{D_1}^2,m_{\phi_1}^2)
\lambda(s,m_{D_2}^2,m_{\phi_2}^2)}\,,
\label{eq.t}
\end{eqnarray}
with $\lambda(a,b,c)=a^2+b^2+c^2-2ab-2bc-2ac$ the K\"all\'en function. 
The $S$-wave amplitude can be obtained by taking $J=0$ in Eq.~\eqref{eq.pwv}. 
In later discussions the subscript $J$ in the partial wave amplitude 
$\mathcal{V}_{J,\,D_1\phi_1\to D_2\phi_2}^{(S,I)}(s)$ will be omitted for simplicity.

The nonperturbative strong interactions of the pNGBs and the ground-state charmed mesons, which manifest
themselves in the emergence of bound states or resonances, can be accounted for by restoring unitarity and
the analytical properties associated with the unitarity cut of the perturbative partial-wave amplitudes in
Eq.~\eqref{eq.pwv}. In this work we use the unitarization approach that has been widely used to discuss the
pNGBs and charmed mesons scattering in  Refs.~\cite{Guo:2006fu,Guo:2009ct,Cleven:2010aw,Wang:2012bu,Altenbuchinger:2013vwa}.
The unitarized amplitude for the two-body scattering process takes the form~\cite{Oller:1998zr,Oller:2000fj} 
\begin{eqnarray} \label{eq.defut}
 T(s) = \big[ 1 - \mV(s)\cdot G(s) \big]^{-1}\cdot \mV(s)\,,
\end{eqnarray}
where $\mV(s)$ denotes the partial-wave amplitude in Eq.~\eqref{eq.pwv} and for simplicity 
both the superscripts and subscripts are omitted. By construction, the $G(s)$ function includes the two-body
unitarity/right-hand cut and it can be given by the loop function 
\begin{eqnarray}\label{eq.defg}
G(s)=i\int\frac{{\rm d}^4q}{(2\pi)^4}
\frac{1}{(q^2-m_{1}^2+i\epsilon)[(P-q)^2-m_{2}^2+i\epsilon ]}\ ,\qquad
s\equiv P^2\ \,.
\end{eqnarray}
One can use a once-subtracted dispersion relation or dimensional regularization 
by replacing the divergence by a constant to calculate the explicit form of the $G(s)$ function,
which reads ~\cite{Oller:1998zr} 
\begin{eqnarray}\label{eq.gfunc}
G(s)^{\rm DR} \al=\al\frac{1}{16\pi^2}\bigg\{{a}(\mu)+\ln\frac{m_{1}^2}{\mu^2}
+\frac{s-m_{1}^2+m_{2}^2}{2s}\ln\frac{m_{2}^2}{m_{1}^2}\nonumber\\
\al\al+\frac{\sigma}{2s}\big[\ln(s-m_{2}^2+m_{1}^2+\sigma)-\ln(-s+m_{2}^2-m_{1}^2+\sigma)\nonumber\\
\al\al+\ln(s+m_{2}^2-m_{1}^2+\sigma)-\ln(-s-m_{2}^2+m_{1}^2+\sigma)\big]\bigg\}\ ,
\end{eqnarray}
where 
\begin{eqnarray}\label{eq.defsigma}
\sigma=\sqrt{\lambda(s,m_1^2,m_2^2)}\,,
\end{eqnarray}
and $\mu$ is the regularization scale. The superscript DR in Eq.~\eqref{eq.gfunc} stresses that the $G(s)$
function in this equation corresponds to the form obtained in dimensional regularization. The function $G(s)^{\rm DR}$
does not depend on the regularization scale $\mu$, since the explicit $\mu$ dependence in Eq.~\eqref{eq.gfunc} is
canceled by that from the subtraction constant $a(\mu)$. In later discussion we take $\mu=1$~GeV in order to allow
for a  comparison with the  previous
works~\cite{Guo:2009ct,Liu:2012zya,Yao:2015qia,Altenbuchinger:2013vwa,Wang:2012bu,Guo:2015dha,Albaladejo:2016lbb}.

The unitarized partial-wave amplitude in Eq.~\eqref{eq.defut} can be easily extended to  coupled-channel scattering,
where one should promote $\mV(s)$ and $G(s)$ to $n \times n$ matrices in case of $n$ channels. The matrix elements
for $\mV(s)$ are given by Eq.~\eqref{eq.pwv}. $G(s)$ becomes a diagonal matrix, with its diagonal elements given
by Eq.~\eqref{eq.gfunc} with the masses $m_1$ and $m_2$ in question. For easy comparison, we follow the previous
works~\cite{Liu:2012zya,Guo:2015dha} for the convention of scattering the length. The $S$-wave scattering length
is related to the unitarized chiral amplitude in Eq.~\eqref{eq.defut} through 
\begin{eqnarray}
 a_{D\phi\to D\phi}=-\frac{1}{8\pi(m_D+m_\phi)}T_{D\phi\to D\phi}(s_{\rm thr}),\qquad s_{\rm thr}=(m_D+m_\phi)^2\,,
\end{eqnarray}
where the superscripts for isospin and strangeness and the subscript for $J=0$ are omitted for simplicity.

\section{ Chiral amplitudes in the finite volume }\label{sect.uniandfv} 

One of the main novelties in this work is to fully exploit the rich 
finite-volume spectra from the lattice simulations given in Ref.~\cite{Moir:2016srx}, 
in order to constrain the unitarized chiral amplitudes. In order to do so, we use the 
method proposed in Refs.~\cite{Doring:2011vk,Doring:2012eu} to introduce the finite-volume
effects into the unitarized chiral amplitudes. As it was demonstrated in Ref.~\cite{Guo:2016zep},
this framework is quite  efficient to fit the lattice finite-volume spectra for the coupled-channel 
scattering of $\pi\eta, K\bar{K}$ and $\pi\eta'$. In this work, we use the
same method to study the coupled-channel $D\pi, D\eta$ and $D_s\bar{K}$ system. 

Below, we briefly describe the method.
The loop function $G(s)$ in Eq.~\eqref{eq.defg} is ultraviolet
divergent and needs to be regularized. One way to do this is to perform 
the integral with the three-momentum cutoff $q_{\rm max}$. 
After integrating over the variable $q^0$ analytically, one gets
\begin{eqnarray}\label{eq.defg3d} 
 G(s)^{\rm cutoff}= \int^{|\vec{q}|<q_{\rm max}} \frac{{\rm d}^3 \vec{q}}{(2\pi)^3} \, I(|\vec{q}|) \,,
\end{eqnarray}
where 
\begin{eqnarray}
I(|\vec{q}|) &=& \frac{w_1+w_2}{2w_1 w_2 \,[E^2-(w_1+w_2)^2]}\,, \nonumber \\
w_i &=&\sqrt{|\vec{q}|^2+m_i^2} \,,  \quad s=E^2 \,. 
\end{eqnarray} 
To obtain the above results when integrating out $q^0$, it is convenient to choose the CM frame, 
by taking the total four-momentum $P^\mu$ of the two-particle system as $(P^0,\vec{P}=0)$. 
Since the $G(s)$ function in the infinite volume, i.e. Eqs.~\eqref{eq.defg} or \eqref{eq.defg3d}, is a Lorentz scalar, 
its final expression is the same in different frames. However due to the breaking of Lorentz invariance in the finite volume, 
one should distinguish the finite-volume quantities defined in different frames. 
The quantities in the CM frame will be denoted with an asterisk in the following.

The finite-volume effects are introduced into the unitarized chiral amplitudes 
by discretizing the above three-momentum integral, defining the loop function. 
The allowed momenta $\vec{q}^{\,*}$ in the cubic box of length $L$ with periodic 
boundary conditions take the discrete values 
\begin{eqnarray}
 \vec{q}^{\,*}= \frac{2\pi}{L} \vec{n}, \quad  \vec{n} \in \mathbb{Z}^3 \,.
\end{eqnarray}
The three-momentum integral in Eq.~\eqref{eq.defg3d} should be replaced by 
the sum of the allowed momenta. Hence, the finite-volume loop function reads 
\begin{eqnarray}\label{eq.gtilde}
 \widetilde{G}= \frac{1}{L^3} \sum_{\vec{n}}^{|\vec{q}^{\,*}|<q_{\rm max}} I(|\vec{q}^{\,*}|)\,.
\end{eqnarray}
Here we introduce a tilde on top of a symbol to distinguish it from 
the same quantity in the infinite volume.

The finite-volume correction $\Delta G$ in the CM frame to the loop function $G(s)$ is then 
given by 
\begin{eqnarray}\label{eq.deltag}
 \Delta G &=& \widetilde{G} - G^{\rm cutoff} \nonumber \\
  & = & \frac{1}{L^3} \sum_{\vec{n}}^{|\vec{q}^{\,*}|<q_{\rm max}} I(|\vec{q}^{\,*}|) -   \int^{|\vec{q}|<q_{\rm max}} 
\frac{{\rm d}^3 \vec{q}}{(2\pi)^3} I(|\vec{q}|)\,. 
\end{eqnarray}
It should be stressed that, as $L\to\infty$, the quantity $\Delta G$ is 
independent of the three-momentum cutoff due to the cancellation of the 
$q_{\rm max}$-dependences of the two terms in this equation and, up to the terms that
vanish exponentially at large $L$, can be related to the pertinent
L\"uscher zeta-function. In practice,
for finite $L$,  it was verified numerically 
(see Ref.~\cite{Guo:2016zep})
that the cutoff dependence of $\Delta G$ is indeed rather weak.
The final expression of the function $G(s)$, used in our finite-volume
analysis, takes the form  
\begin{eqnarray}\label{eq.gfuncfvdr}
 \widetilde{G}^{\rm DR}= G^{\rm DR} + \Delta G \,,
\end{eqnarray}
where $G^{\rm DR}$ and $\Delta G$ are explicitly given in Eqs.~\eqref{eq.gfunc} 
and \eqref{eq.deltag}, respectively.

As mentioned previously, although the loop function $G^{\rm DR}(s)$ 
in the infinite volume is Lorentz invariant, the corresponding finite volume expression in 
Eq.~\eqref{eq.gfuncfvdr} is not Lorentz invariant any more. As a result, one has to 
explicitly work out the different expressions for the loop functions 
in different frames, which are considered in 
Refs.~\cite{Doring:2012eu,Gockeler:2012yj,Roca:2012rx,Fu:2011xz}. Here 
we recapitulate the main results to set up the notation.

For the two-body system, moving with the four-momentum 
$P^\mu=(P^0, \vec{P})$, the CM energy squared is $s=E^2=(P^0)^2-|\vec{P}|^2$ 
and the three-momenta of the particles in the moving frame 
are $\vec{q_1}$ and $\vec{q}_2$, respectively,
with $\vec{q}_1+ \vec{q}_2=\vec{P}$. 
The corresponding three-momenta in the CM frame are denoted 
by $\vec{q_1}^{\,*}$ and $\vec{q}_2^{\,*}$, respectively, with 
$\vec{q}_1^{\,*}=- \vec{q}_2^{\,*}=\vec{q}^*$. By performing the 
standard Lorentz boost, one obtains 
\begin{eqnarray}\label{eq.lrtr2}
 \vec{q}_i^{\,*}= \vec{q}_i + \bigg[ \bigg(\frac{E}{P^0} - 1\bigg)
\frac{\vec{q}_i\cdot \vec{P}}{|\vec{P}|^2} 
- \frac{q_i^{*\,0}}{P^0} \bigg]\vec{P}\,, 
\end{eqnarray} 
where the on-shell energies $q_{1}^{\,*\,0}$ and $q_{2}^{\,*\,0}$ take the form  
\begin{eqnarray}
 q_1^{\,*\,0}= \frac{E^2+m_1^2-m_2^2}{2E} \,, \quad  q_2^{\,*\,0}= \frac{E^2+m_2^2-m_1^2}{2E} \,.
\end{eqnarray}
With these definitions, the finite-volume loop function in the 
moving frame reads~\cite{Doring:2012eu}
\begin{eqnarray}\label{eq.gfuncfvmv}
\widetilde{G}^{\rm MV}= \frac{E}{P^0 L^3}  \sum_{\vec{q}}^{|\vec{q}^{\,*}|<q_{\rm max}} I(|\vec{q}^{\,*}(\vec{q})|)\,,
\end{eqnarray} 
with 
\begin{eqnarray}
\vec{q}=\frac{2\pi}{L}\vec{n}\,, \quad \vec{P}=\frac{2\pi}{L}\vec{N}\,,  \quad  (\vec{n}, \vec{N}) \in \mathbb{Z}^3\,. \label{eq.pfv}
\end{eqnarray} 
It is obvious that the expression in 
Eq.~\eqref{eq.gfuncfvmv} in the moving frame recovers the formula 
of Eq.~\eqref{eq.gtilde}, defined in the CM frame with $\vec{P}=0$. 
In analogy with Eq.~\eqref{eq.gfuncfvdr} in the CM frame, 
the final expression for the loop function used in the moving frame takes the 
form 
\begin{eqnarray}\label{eq.gfuncfvdrmv}
 \widetilde{G}^{\rm DR, MV}= G^{\rm DR} + \Delta G^{\rm MV}  \,,
\end{eqnarray}
where
\begin{equation} 
\Delta G^{\rm MV} =   \widetilde{G}^{\rm MV} - G^{\rm cutoff}\,,
\end{equation}
with $G^{\rm DR}$, $\widetilde{G}^{\rm MV}$ and $G^{\rm cutoff}$ given in 
Eqs.~\eqref{eq.gfunc}, \eqref{eq.gfuncfvmv} and \eqref{eq.defg3d}, respectively\footnote{The equation~\eqref{eq.defg3d} is written down in the rest frame. The corresponding expression in the arbitrary moving frame is the same, only the energy $E$ in the denominator is replaced by $\sqrt{s}$. }.  
In order to account for the higher partial waves in the determination of the finite-volume energy levels, the generalized $G(s)$ functions are introduced~\cite{Doring:2012eu} 
\begin{eqnarray}\label{eq.defghp2l}
  \widetilde{G}^{\rm MV}_{\ell m, \ell' m'}= \frac{4\pi}{L^3} \frac{E}{P^0} \sum_{\vec{n}}^{|\vec{q\,}^*|<q_{\rm max}}
  \left( \frac{|\vec{q\,}^*|}{|\vec{q\,}^{\rm on*}|}\right)^{k} Y_{\ell m}^{*}(\hat{q}^{*}) Y_{\ell' m'}(\hat{q}^{*})\,I(|\vec{q\,}^*|)\,, 
\end{eqnarray}
where $|\vec{q}^{\rm on*}|$ denotes the on-shell value for $|\vec{q\,}^*|$, $\hat{q}^{*}=\vec{q\,}^{*}/|\vec{q\,}^*|$, $k=0~(1)$ for $\ell+\ell'=$ even~(odd), and the  $Y_{\ell m}$ denote the spherical harmonics functions with the normalization
\begin{eqnarray}
\int_{0}^{2\pi} d\phi \int_{0}^{\pi} \sin\theta d\theta Y_{\ell m}(\theta,\phi) Y_{\ell' m'}^{*}(\theta,\phi) = \delta_{\ell \ell'} \delta_{m m'}\,.
\end{eqnarray}
One can establish the relation of $\widetilde{G}^{\rm  MV}_{\ell m, \ell' m'}$ in Eq.~\eqref{eq.defghp2l} with $\mathcal{M}_{\ell m, \ell' m'}$ (the linear combination of the L{\"u}scher zeta functions) in Eq.~(39) of Ref.~\cite{Gockeler:2012yj}. See also Refs.~\cite{Rummukainen:1995vs,Leskovec:2012gb} for further details on $\mathcal{M}_{\ell m, \ell' m'}$. 

Further, it is convenient to introduce the quantity
\begin{eqnarray}\label{eq.defghp1l}
\widetilde{G}^{\rm MV}_{\ell m }= \sqrt{\frac{4\pi}{2\ell + 1}}\frac{1}{L^3} \frac{E}{P^0}  \sum_{\vec{n}}^{|\vec{q\,}^*|<q_{\rm max}} \left( \frac{|\vec{q}^*|}{|\vec{q}^{\rm on*}|}\right)^{\ell}  Y_{\ell m}(\hat{q}^{*}) \, I(|\vec{q\,}^*|)\,,
\end{eqnarray}
which, up to the exponentially suppressed terms, 
is related to the quantity $w_{\ell m}$,
defined in Eq.~(40) of Ref.~\cite{Gockeler:2012yj}, through
\begin{eqnarray}
 \widetilde{G}^{\rm  MV}_{\ell m } = -\frac{|\vec{q\,}^{\rm on*}|}{8\pi E} w_{\ell m}\,.
\end{eqnarray}
In analogy to Eq.~\eqref{eq.gfuncfvdrmv}, 
we define
\begin{eqnarray}\label{eq.gfuncfvdrmvhp1l}
 \widetilde{G}^{\rm DR, MV}_{\ell m}= G^{\rm DR}\delta_{\ell\,0}\delta_{m\,0} +  \Delta G^{\rm MV}_{\ell m}  \,,
\end{eqnarray}
where
\begin{equation} 
\Delta G^{\rm MV}_{\ell m} =   \widetilde{G}^{\rm MV}_{\ell m} - G^{\rm cutoff}\delta_{\ell\,0}\delta_{m\,0}\,,
\end{equation}
with $G^{\rm DR}$, $\widetilde{G}^{\rm MV}_{\ell m}$ and $G^{\rm cutoff}$ given in 
Eqs.~\eqref{eq.gfunc}, \eqref{eq.defghp1l} and \eqref{eq.defg3d}, respectively. It is easy to show that by  taking $\ell=0, m=0$ and $\vec{P}=0$, Eq.~\eqref{eq.gfuncfvdrmvhp1l} reduces to the CM formula of Eq.~\eqref{eq.gfuncfvdr}, as it should be. In order to simplify the notation, we will denote $\widetilde{G}^{\rm DR, MV}_{\ell m}$ in  Eq.~\eqref{eq.gfuncfvdrmvhp1l} by  
$\widetilde{G}_{\ell m}$ in the following. 

Due to the rotational invariance, different partial waves do not mix
 in the infinite volume. However, this feature is lost in a finite volume,
and the different partial-wave amplitudes $\mathcal{V}_{J}(s)$ in 
Eq.~\eqref{eq.pwv} get mixed. A more subtle issue is that the mixing patterns 
of the partial-wave amplitudes vary in different moving frames. 
In the following, 
we shall retain only the $S$- and $P$-wave amplitudes of the 
$D\pi, D\eta$ and $D_s\bar{K}$ system, which should be a reasonable 
approximation up to the $D_s\bar{K}$ threshold energy  
region~\cite{Moir:2016srx}. 

The projection of the two-body quantization condition onto the irreducible
representations of the different little groups of the octahedral group
$O_h$, corresponding to the different moving frames, has been carried out in
all details in Ref.~\cite{Gockeler:2012yj}. In this paper, we wish to adapt 
these results for the case of the unitarized ChPT in a finite volume.
The pertinent formulas can be directly read off
from Ref.~\cite{Gockeler:2012yj}, replacing $w_{\ell m}$ by  the
quantity $\widetilde{G}_{\ell m}$ introduced above, and keeping track of 
the normalization factors.
Of course, in the present work we consider the coupled-channel scattering, 
but this does not change the symmetry properties of the equation as, 
simply, in addition, the amplitudes become matrices in the channel space.

Below, we display the explicit equations in different frames.
In the CM frame, there is no mixing between $S$- and $P$-wave scattering amplitudes. For the $S$-wave in the $A_{1}^{+}$ irreducible representation, the finite-volume energy levels are given by the solutions of the equation~\cite{Doring:2012eu,Gockeler:2012yj} 
\begin{eqnarray}\label{eq.detmvs} 
 \det[I- \mathcal{V}_{0}(s)\cdot \widetilde{G}_{00}] = 0\,,
\end{eqnarray}
where $I$ is  the unit matrix and the matrix elements of $\mathcal{V}_{0}(s)$ and $\widetilde{G}_{00}$ can be calculated via  Eqs.~\eqref{eq.pwv} and ~\eqref{eq.gfuncfvdrmvhp1l}, respectively. 

Further, according to Ref.~\cite{Moir:2016srx}, there exists a bound state in the $P$-wave $D\pi$ scattering. In order to consider the contribution of the
$P$-wave to the energy levels, we use a simple ansatz to include the bound state
\begin{eqnarray}\label{eq.v1}
\mathcal{V}_{1}(s)= \frac{g_V^2[s-(m_D+m_\pi)^2][s-(m_D-m_\pi)^2]}{4s(s-m_{D^*}^2)}\,,
\end{eqnarray}
where the superscripts $(0,\frac{1}{2})$ of $\mathcal{V}_{1}$ are omitted for later convenience, and $g_V$ and $m_{D^*}$ shall be adjusted to reproduce the lattice energy levels.  For the $P$ wave in the $T_{1}^{-}$ irreducible representation, the finite-volume energy levels are determined by~\cite{Doring:2012eu}
\begin{eqnarray}\label{eq.detmvp} 
 \det[I- \mathcal{V}_{1}(s)\cdot \widetilde{G}_{00} ] = 0\,,
\end{eqnarray}
where $\mathcal{V}_{1}(s)$ and $\widetilde{G}_{\ell m}$ are given in Eqs.~\eqref{eq.v1} and \eqref{eq.gfuncfvdrmvhp1l}, respectively. We mention that the determinant in the above equation is in fact trivial,  since the single-channel approximation is used for the $P$-wave scattering.

In the moving frame with the total three-momentum $\vec{P}=(2\pi/L)\vec{N}$, the $S$- and $P$-wave amplitudes will get mixed. For the moving frame with $\vec{N}=(0,0,1)$, the equation to determine the discrete energy levels in the irreducible representation $A_1$ is
\begin{eqnarray}\label{eq.det001a1}
 \det[I - \mathcal{V}_{0,1}\cdot\mathcal{M}_{0,1}^{A_1} ] = 0\,,
\end{eqnarray}
where
\begin{eqnarray}\label{eq.v01}
\mathcal{V}_{0,1}= \bma
\mathcal{V}_0 & 0 \\
0  &  \mathcal{V}_1 
\ema \,,
\end{eqnarray}
\begin{eqnarray}\label{eq.m001a1}
\mathcal{M}_{0,1}^{A_1}= \bma
\widetilde{G}_{00} & i\sqrt{3}\widetilde{G}_{10} \\
-i\sqrt{3}\widetilde{G}_{10} & \widetilde{G}_{00} + 2\widetilde{G}_{20} 
\ema \,.
\end{eqnarray}
Here, $\mathcal{V}_{0}$, $\mathcal{V}_{1}$ and $\widetilde{G}_{\ell m}$ should be understood as matrices in the scattering-channel space. To be more specific, the $S$-wave $\mathcal{V}_{0}$ corresponds to a $3\times 3$ matrix, spanned by the $D\pi, D\eta$ and $D_s\bar{K}$ channels. For $\mathcal{V}_{1}$ it is an ordinary function, since the single-channel approximation is taken for the $P$ wave. As a result, the $4\times 4$ matrix of $\mathcal{V}_{0,1}\cdot\mathcal{M}_{0,1}^{A_1}$ in Eq.~\eqref{eq.det001a1} is given by
\begin{eqnarray}
\mathcal{V}_{0,1}\cdot\mathcal{M}_{0,1}^{A_1} =   \bma
      \mathcal{V}_{0,11}\widetilde{G}_{00,1} & \mathcal{V}_{0,12}\widetilde{G}_{00,2} & \mathcal{V}_{0,13}\widetilde{G}_{00,3} &  i\sqrt{3}\mathcal{V}_{0,11}\widetilde{G}_{10,1}\\
     \mathcal{V}_{0,21}\widetilde{G}_{00,1} &  \mathcal{V}_{0,22}\widetilde{G}_{00,2} & \mathcal{V}_{0,23}\widetilde{G}_{00,3} &  i\sqrt{3}\mathcal{V}_{0,21}\widetilde{G}_{10,1}\\
     \mathcal{V}_{0,31}\widetilde{G}_{00,1} & \mathcal{V}_{0,32}\widetilde{G}_{00,2} &  \mathcal{V}_{0,33}\widetilde{G}_{00,3} &  i\sqrt{3}\mathcal{V}_{0,31}\widetilde{G}_{10,1}\\
     -i\sqrt{3}\mathcal{V}_{1}\widetilde{G}_{00,1} & 0 & 0 &  \mathcal{V}_{1}(\widetilde{G}_{00,1} + 2\widetilde{G}_{20,1})\\
     \ema\,, 
\end{eqnarray}
where $i$ and $j$ in the subscripts of $\mathcal{V}_{0,ij}$ and $\widetilde{G}_{\ell m, i}$ are the channel indices. The $D\pi, D\eta$ and $D_s\bar{K}$ channels are labeled by 1, 2,  and 3, respectively. 

For other moving frames, the corresponding equations to determine the discrete energy levels for the irreducible representation $A_1$ can be obtained by replacing the $\mathcal{M}_{0,1}^{A_1}$ in  Eq.~\eqref{eq.det001a1} with the proper ones, which are given in Ref.~\cite{Gockeler:2012yj}. We quote the explicit results below for completeness. 
For $\vec{N}=(1,1,0)$, it is
\begin{eqnarray}\label{eq.110a1}
\mathcal{M}_{0,1}^{A_1}= \bma
\widetilde{G}_{00} & -\sqrt{6}(1-i)\,{\rm Re}[\widetilde{G}_{11}] \\
-\sqrt{6}(1+i)\,{\rm Re}[\widetilde{G}_{11}] & \widetilde{G}_{00}-\widetilde{G}_{20}-i\sqrt{6}\widetilde{G}_{22} 
\ema \,.
\end{eqnarray}
For $\vec{N}=(1,1,1)$, it is
\begin{eqnarray}\label{eq.m111a1}
\mathcal{M}_{0,1}^{A_1}= \bma
\widetilde{G}_{00} & \frac{3}{\sqrt{2}}(1-i)\widetilde{G}_{10} \\
\frac{3}{\sqrt{2}}(1+i)\widetilde{G}_{10} & \widetilde{G}_{00}-i2\sqrt{6}\widetilde{G}_{22} 
\ema \,.
\end{eqnarray}
The partial-wave scattering amplitudes $\mathcal{V}_0$ and $\mathcal{V}_1$ are the same as those in Eq.~\eqref{eq.det001a1} for different moving frames and different irreducible representations.

For the irreducible representations $E$  
when $\vec{N}=(0,0,1)$, $B_1$ and $B_2$ when $\vec{N}=(1,1,0)$ and $E$ when $\vec{N}=(1,1,1)$, the $S$-wave amplitudes are decoupled and only the $P$ wave enters. The general equation to determine the discrete energy levels is given by the solution of
\begin{eqnarray}\label{eq.detmvpgeneral} 
\det[I- \mathcal{V}_{1}(s)\cdot \mathcal{M}_1 ] = 0\,.
\end{eqnarray}
According to Ref.~\cite{Gockeler:2012yj}, $\mathcal{M}_1$ takes different forms for different representations. For the irreducible presentation $E$ when $\vec{N}=(0,0,1)$, it reads 
\begin{eqnarray}\label{eq.m1001e2} 
\mathcal{M}_1  = \widetilde{G}_{00} - \widetilde{G}_{20}\,. 
\end{eqnarray}
For the irreducible presentation $B_1$ when $\vec{N}=(1,1,0)$, it reads 
\begin{eqnarray}\label{eq.m1110b1} 
\mathcal{M}_1  = \widetilde{G}_{00} + 2\widetilde{G}_{20}\,. 
\end{eqnarray}
For the irreducible presentation $B_2$ when $\vec{N}=(1,1,0)$, it reads
\begin{eqnarray}\label{eq.m1110b2} 
\mathcal{M}_1  = \widetilde{G}_{00} - \widetilde{G}_{20} + i\sqrt{6}\widetilde{G}_{22}\,. 
\end{eqnarray}
For the irreducible presentation $E$ 
when $\vec{N}=(1,1,1)$, it reads
\begin{eqnarray}\label{eq.m1111e2} 
\mathcal{M}_1  = \widetilde{G}_{00} + i\sqrt{6}\widetilde{G}_{22}\,. 
\end{eqnarray}
The partial-wave amplitude $\mathcal{V}_1(s)$ in different representations takes the same expression in Eq.~\eqref{eq.v1}.  

All formulas, which are relevant for further discussions, were listed above.
The formulas for other irreducible representations will not be explicitly 
given here. We refer to Ref.~\cite{Gockeler:2012yj} for further details.

\section{Fits to the finite-volume spectra and scattering lengths from lattice calculations}\label{sect.fit} 

In order to precisely determine the scattering amplitudes of the charmed and light pseudoscalar mesons, we perform global fits to the discrete finite-volume spectra and the scattering lengths from several lattice  calculations~\cite{Liu:2012zya,Mohler:2013rwa,Lang:2014yfa,Moir:2016srx}. To be more specific, we include the finite-volume spectra,
which were used in Ref.~\cite{Moir:2016srx} to study the $S$- and $P$-wave $D\pi, D\eta$ and $D_s\bar{K}$ coupled-channel scattering with $I=1/2$, and which amount to 47 data points in total (38 data points below the $D_s\bar{K}$ threshold)~\footnote{We greatly appreciate the Hadron Spectrum Collaboration (HSC) to kindly provide us the lattice data with correlation coefficients.}. In addition, the elastic scattering lengths obtained with $m_\pi<600$~MeV from Ref.~\cite{Liu:2012zya}, which amount to 15 data points, are incorporated in our fits. The 2+1 flavor lattice calculation of the $DK$ scattering length and the two energy levels well below the $D_s\eta$ threshold with $(S,I)=(1,0)$ from Refs.~\cite{Mohler:2013rwa,Lang:2014yfa} are also considered in the global fits, which amount to 3 additional data points. In the previous reference, the $DK$ scattering lengths with relatively small statistical uncertainties are obtained from the lowest two energy levels within the effective-range-expansion framework, which provides an efficient method to determine quantities at thresholds. We have also tried to only fit the lowest two energy levels, which turn out to be quite close to the present results.

Regarding the finite-volume spectra from the $S$- and $P$-wave coupled $D\pi, D\eta$ and $D_s\bar{K}$ scattering with $I=1/2$, in one fit strategy we use exactly the same 47 data points as those in Ref.~\cite{Moir:2016srx} to determine the scattering amplitudes, which amounts to 65 data points in total. The present study is based on the chiral amplitudes with NLO local interactions and all the bound states or resonances are generated through the unitarization procedure. Therefore it is not expected that we can reliably describe the strong dynamics well above the scattering threshold. To make an estimate of the systematic error, in another fit strategy we only include the 38 points below the $D_s\bar{K}$ threshold among the overall 47 data from Ref.~\cite{Moir:2016srx} in our study, which amounts to 56 data points in total.

Before going to the details of the fits, we comment on the value of pion decay constant $F_\pi$ appearing in Eq.~\eqref{eq.v}. One approach is to use the physical value $F_\pi=92.1$~MeV when fitting the lattice results calculated at unphysical pion masses, as done in Refs.~\cite{Liu:2012zya,Guo:2015dha}. Another approach is to use the unphysical $F_{\pi}$ values at the corresponding unphysical pion masses. For the lattice data in Ref.~\cite{Liu:2012zya}, the values of $F_{\pi}$ have been calculated in Ref.~\cite{WalkerLoud:2008bp} and we take the values therein. The $F_{\pi}$ value for the lattice used in Ref.~\cite{Moir:2016srx} has not been given from lattice calculation. We take the chiral extrapolated value $F_\pi=105.9$~MeV  determined in our previous work~\cite{Guo:2016zep}. The physical $F_{\pi}$ value is used to study the lattice data of  Refs.~\cite{Mohler:2013rwa,Lang:2014yfa}, since the pion mass used in the lattice calculation is quite close to the physical value. From the chiral power counting point of view, there is no preference as to which approach to use up to the order considered in this work. The discrepancy resulting from the two approaches can be considered as a systematic uncertainty. One may also think of introducing different pNGB decay constants, such as $F_\pi$ and $F_K$, to different channels. In practice we do not expect  this effect as important as the differences between the physical and unphysical values of $F_\pi$ when performing the chiral extrapolation in next section. This has been explicitly verified in Ref.~\cite{Guo:2016zep}, where we have carried out  the calculation to study the effects by using different pNGB decay constants in different channels, which indeed turn out to be small. One of the reasons is that the shift of $F_K$ when extrapolating $m_\pi$ from $391$~MeV~\cite{Moir:2016srx} to the physical value is very moderate, which is estimated to be from $115$~MeV to $110$~MeV in Ref.~\cite{Guo:2016zep}. As a result, we do not introduce another type of fit to distinguish different pNGB decay constants in this work.

Four different types of fits are performed in our study by using different data sets and different $F_{\pi}$ values as discussed above. In the following we denote the four fits by Fit-1A, Fit-1B, Fit-2A and Fit-2B. In the notations, 1 and 2 stand for the two different data sets. 1 is for the 56 data points and 2 the 65 data points. A and B stand for the two different choices of the $F_{\pi}$ values. A means using the unphysical $F_{\pi}$ values for the lattice data at the unphysical pion masses, while B means using physical $F_{\pi}$ for all lattice data.

When only fitting the elastic scattering lengths from the lattice simulations in Refs.~\cite{Liu:2012zya,Mohler:2013rwa,Lang:2014yfa}, it was found that for all the channels one common subtraction constant, defined in Eq.~\eqref{eq.gfunc}, is able to satisfactorily reproduce the lattice  results~\cite{Liu:2012zya,Guo:2015dha,Yao:2015qia,Du:2017ttu,Altenbuchinger:2013vwa,Wang:2012bu}. The same value of the subtraction constant $a(\mu)$ determined from the elastic channels~\cite{Liu:2012zya} was also used in the coupled-channel $D\pi, D\eta$ and $D_s\bar{K}$ $S$-wave scattering to predict resonance poles of $\dd$ in Ref.~\cite{Albaladejo:2016lbb}. 
However, it is found that one common subtraction constant is not sufficient any more when simultaneously including the finite-volume spectra~\cite{Moir:2016srx} and the elastic scattering lengths~\cite{Liu:2012zya,Mohler:2013rwa,Lang:2014yfa} in the global fits. For the $S$-wave coupled-channel $D\pi, D\eta$ and $D_s\bar{K}$ scattering with $I=1/2$, two subtraction constants ${a}_{D\pi}^{0,1/2}$ and ${a}_{D\eta}^{0,1/2}$ are needed to reasonably describe the finite-volume spectra. We fix the subtraction constant in the $D_s\bar{K}$ channel to be the same as in the $D\eta$ channel since, as seen {\it a posteriori,} the fit quality does not improve in general by introducing a free $D_s\bar{K}$ subtraction constant. Further, in our study we use the single-channel formula to fit the two energy levels well below the $D_s\eta$ threshold in analogy to the lattice study of the $I=0$ $DK$ scattering~\cite{Mohler:2013rwa,Lang:2014yfa}, which did not consider the coupling of $D_s\eta$ channel. We find that a common subtraction constant ${a}_{DK}^{1,0}$ for both the $DK$ and $D_s\eta$ channels is able to well reproduce the lattice scattering length.  For all other channels listed in Table~\ref{tab:ci}, a common subtraction constant ${a}_{EC}$ is used and we find it is sufficient to describe the scattering lengths given by the lattice calculation~\cite{Liu:2012zya}.

The elastic $P$-wave scattering amplitude in Eq.~\eqref{eq.v1} is incorporated in our study to describe the finite-volume spectra in Ref.~\cite{Moir:2016srx}. According to the energy levels in Fig.~3 of that reference, clearly there is a bound state well below the $D\pi$ threshold in the $P$-wave amplitude. Furthermore, the similarities of the lowest levels in Figs.~2 and 3 in Ref.~\cite{Moir:2016srx}, indicate that the $S$- and $P$-wave mixing effects are weak, which also justifies the elastic approximation of the $P$-wave amplitude. The lowest energy levels in Fig.~3 of Ref.~\cite{Moir:2016srx} are dominated by the $P$-wave  amplitude, which determines $m_{D^*}=2009$~MeV in Eq.~\eqref{eq.v1}. For the coupling $g_V$, we find that the fits are rather insensitive to its value. Therefore we fix $m_{D^*}=2009$~MeV and $g_V=3$ in the following discussions. It is verified that the fits are barely affected by varying $g_V$ in a wide range from $0.5$ to $5$. The subtraction constant in the $P$-wave amplitude is fixed to be equal to the value in the $S$-wave case.

There are six LECs $h_{i=0,\ldots,5}$ in the NLO scattering amplitude. The values of $h_0$ and $h_1$ can be fixed to be $h_0=0.033$ and $h_1=0.43$ from the masses of $D$ and $D_s$~\cite{Guo:2015dha}, comparing with the slightly different values $h_0=0.014$ and $h_1=0.42$ used in Ref.~\cite{Liu:2012zya}. We still have 8 parameters, i.e. the remaining four LECs $h_{i=2, 3, 4, 5}$ and the four subtraction constants ${a}_{D\pi}^{0,1/2}$, ${a}_{D\eta}^{0,1/2}$, ${a}_{DK}^{1,0}$ and ${a}_{EC}$,  which need to be determined from the fits to the lattice data. As has been done in Refs.~\cite{Liu:2012zya,Guo:2015dha}, we redefine the LECs $h_{i=2, 3, 4, 5}$ as follows in order to reduce the correlations in the fits:
\begin{equation}
h_{24}\equiv h_2+h_4^\prime\ ,\quad  h_{35}\equiv h_3+2\,h_5^\prime\ , \quad h_4^\prime\equiv h_4 \hat{M}_D^2\ ,  \quad h_5^\prime\equiv h_5 \hat{M}_D^2\ , 
\end{equation}
where $\hat{M}_D\equiv(M_D^{\rm phys}+M_{D_s}^{\rm phys})/2$. 
Unlike the subtraction constants that each of them can only enter in a specific channel, every single chiral LEC could appear in all the scattering amplitudes. This is another reason that urges us to perform global fits by including the finite-volume energy levels of the coupled-channel $D\pi,D\eta$ and $D_s\bar{K}$ scattering~\cite{Moir:2016srx} as well as the scattering lengths of various channels given in Refs.~\cite{Liu:2012zya,Mohler:2013rwa,Lang:2014yfa}. The values of the parameters from the four types of fits are collected in Table~\ref{tab.fitlecs}. The results from Ref.~\cite{Liu:2012zya} are also presented in the last column for comparison.

\begin{table}[thbp]
\centering
\begin{small}
\begin{tabular}{ c c c c c c }
\hline\hline
 & Fit-1A & Fit-1B   & Fit-2A  & Fit-2B  & Table V~\cite{Liu:2012zya} 
\\ \hline
$h_{24}$       &$-0.50_{-0.11}^{+0.12}$   &$-0.64_{-0.11}^{+0.17}$  &$-0.42_{-0.15}^{+0.18}$  &$-0.14_{-0.14}^{+0.10}$ &$-0.10_{-0.06}^{+0.05}$
\\
$h_{4}^\prime$ & $-1.45_{-0.61}^{+0.68}$ &$-1.30_{-0.68}^{+0.50}$  &$-0.49_{-0.23}^{+0.23}$  &$-0.02_{-0.36}^{+0.34}$ &$-0.32_{-0.34}^{+0.35}$
\\
$h_{35}$       & $0.83_{-0.19}^{+0.13}$  &$0.77_{-0.21}^{+0.14}$   &$0.76_{-0.22}^{+0.16}$  &$0.05_{-0.12}^{+0.16}$  &$0.25_{-0.13}^{+0.13}$
\\
$h_{5}^\prime$ & $0.74_{-0.68}^{+0.78}$  & $0.68_{-0.51}^{+0.56}$  & $-0.49_{-0.17}^{+0.17}$ &$-0.81_{-0.32}^{+0.33}$ &$-1.88_{-0.61}^{+0.63}$
\\
${a}_{D\pi}^{0,1/2}$  & $-1.73_{-0.19}^{+0.21}$ & $-1.45_{-0.14}^{+0.19}$  &$-2.00_{-0.12}^{+0.13}$ &$-1.52_{-0.06}^{+0.07}$ &$-1.88_{-0.09}^{+0.07\,*}$
\\
${a}_{D\eta}^{0,1/2}$ & $-2.68_{-0.19}^{+0.21}$ & $-2.53_{-0.25}^{+0.24}$ & $-2.43_{-0.24}^{+0.21}$ &$-2.02_{-0.10}^{+0.08}$ &$-1.88_{-0.09}^{+0.07\,*}$
\\
${a}_{DK}^{1,0}$     & $-1.58_{-0.22}^{+0.17}$& $-1.62_{-0.18}^{+0.16}$  &$-1.86_{-0.27}^{+0.18}$  &$-1.60_{-0.17}^{+0.11}$ &$-1.88_{-0.09}^{+0.07\,*}$
\\
${a}_{EC}$            & $-2.72_{-0.21}^{+0.20}$ & $-2.69_{-0.20}^{+0.18}$ &$-2.45_{-0.19}^{+0.23}$  &$-1.91_{-0.25}^{+0.18}$ &$-1.88_{-0.09}^{+0.07}$
\\ \hline
$\chi^2/{\rm d.o.f}$ & $116.7/(56-8)$  & $124.1/(56-8)$  & $221.8/(65-8)$  & $215.5/(65-8)$& $1.06$
\\
\hline\hline
\end{tabular}
\end{small}
\caption{\label{tab.fitlecs} Fitting results of the four type of fits. See the text for the details about the four fits. The results from Ref.~\cite{Liu:2012zya} are presented in the last column for comparison. The asterisks for the corresponding subtractions denote that their values are simply imposed to be equal to the fitted elastic channel result ${a}_{EC}$.}
\end{table}

We would like to mention that the correlations between different energy levels within the same volume from Ref.~\cite{Moir:2016srx} are included in our fits. 
If the correlations are neglected, the resulting $\chi^2$ will be greatly reduced, which turn out to be $75.5, 87.6, 134.6$ and $132.0$ for Fit-1A, Fit-1B, Fit-2A and Fit-2B, respectively. Further, three sets of data from different lattice collaborations using rather different ensembles are included in our fits and they may introduce potentially large systematical uncertainties, which are difficult to estimate and hence are not included in this work. This provides another explanation of the somewhat large $\chi^2$ from our fits. In fact, we have tried to only fit the 47 data from the HSC for the coupled $D\pi, D\eta$ and $D_s\bar{K}$ scattering with $I=1/2$. By releasing all the six chiral low energy constants in Eq.~\eqref{nlolag}, it is possible for us to obtain much smaller $\chi^2$ values for the HSC data~\cite{Moir:2016srx}. However the resulting values of $h_0$ and $h_1$, are significantly different from the results of Refs.~\cite{Liu:2012zya,Guo:2015dha}, which are determined by properly reproducing the masses of the grounds-state charmed mesons. Given the fact that the $\chi^2/{\rm d.o.f.}$ for the Fit-2A and Fit-2B are around 4 one could consider the possibility to double the relative errors of the data fitted in order to estimate the precision achieved by employing our parameterization based on unitarized ChPT. By taking into account that the relative errors for the data of Ref.~\cite{Moir:2016srx} range in an interval of around $0.05-0.6\%$, this would imply that we are able to give a fair reproduction of the lattice QCD data at the level of a $0.1-1.2\%$, which indeed is a great achievement for a parameterization based on unitarized $SU(3)$ NLO ChPT. The latter is expected to be affected by errors from higher-order corrections at the level of $[(m_\pi \sim m_K)/1~{\rm GeV}]^3$, i.e. around $6\sim 15\%$ for the unphysically large meson masses used here. One should take into account that by unitarizing ChPT the resulting parameterization is expected to be more precise, particularly if the data reflect the presence of resonances that are properly reproduced with the nonperturbative approach. One possible way to improve the discussions is to generalize the present study to next-to-next-to-leading order~\cite{Yao:2015qia,Du:2017ttu}, which is clearly beyond the scope of this work. As a result, we shall focus on the more constrained fits shown in Table~\ref{tab.fitlecs} in the following discussions.

As shown in Table~\ref{tab.fitlecs}, the $\chi^2$ values resulting from Fit-2A and Fit-2B, which include the finite-volume energy levels above the $D_s\bar{K}$ threshold from Ref.~\cite{Moir:2016srx}, are clearly larger than those from Fit-1A and Fit-1B that only include the finite-volume spectra below the $D_s\bar{K}$ threshold. We further verify that the extra amounts of the $\chi^2$ from Fit-2A and Fit-2B are mainly contributed by the finite-volume energy levels from the $D\pi, D\eta$ and $D_s\bar{K}$ coupled-channel scattering. Comparing the four types of fits with the results in the last column in Table~\ref{tab.fitlecs}, we observe that the parameters from Fit-2B are the closest to the values given in Ref.~\cite{Liu:2012zya}. According to the large $N_C$ arguments, $h_2$ is expected to be $1/N_C$ suppressed comparing with $h_3$. The same expectation is also applied to $h_4$ and $h_5$. In Ref.~\cite{Du:2016tgp}, it provides another useful theoretical criteria to discriminate different parameter sets, which relies on the positive constraints of the scattering amplitudes. If one considers the $N_C$ argument and the positivity bound, it is plausible that Fit-2B is the preferred one comparing with the other three fits in Table~\ref{tab.fitlecs}. 
We also find an additional solution for Fit-2A, which gives similar total $\chi^2$. However, the additional solution gives a worse description of the elastic scattering lengths in  Ref.~\cite{Liu:2012zya} and the energy levels of the $DK$ scattering in  Refs.~\cite{Mohler:2013rwa,Lang:2014yfa} than the other fits in Table~\ref{tab.fitlecs}. In this respect, the other additional solution of Fit-2A is considered to be disfavored and we refrain from discussing the results from that solution.

\begin{figure}[htbp]
\centering
\includegraphics[width=0.95\textwidth]{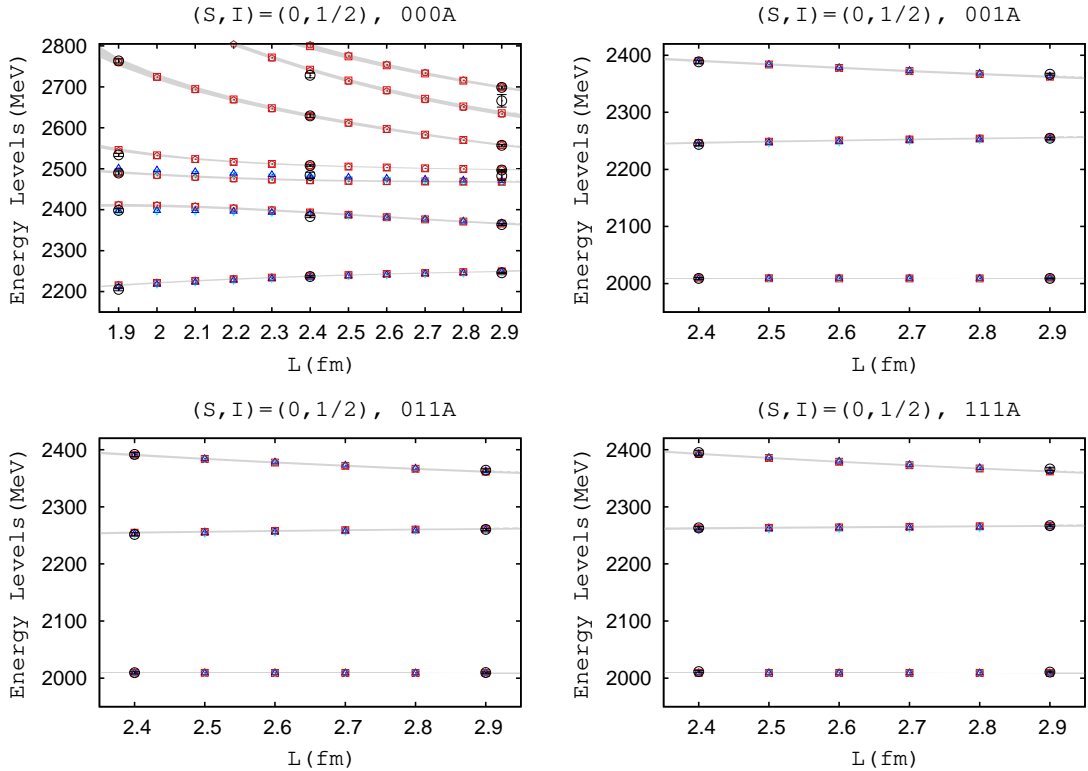} 
\caption{(Color online) Finite-volume energy levels in different frames for the $D\pi, D\eta$ and $D_s\bar{K}$ scattering with $(S,I)=(0,1/2)$. Cyan downward triangles, blue upward triangles and brown pentagons correspond to the central-value results of Fit-1A, Fit-1B and Fit-2A, respectively. Red squares denote the results from Fit-2B and the gray shaded areas denote the corresponding one-sigma error bands. The lattice data are taken from Ref.~\cite{Moir:2016srx}. }\label{fig.lev1}
\end{figure}

\begin{figure}[htbp]
\centering
\includegraphics[width=0.95\textwidth]{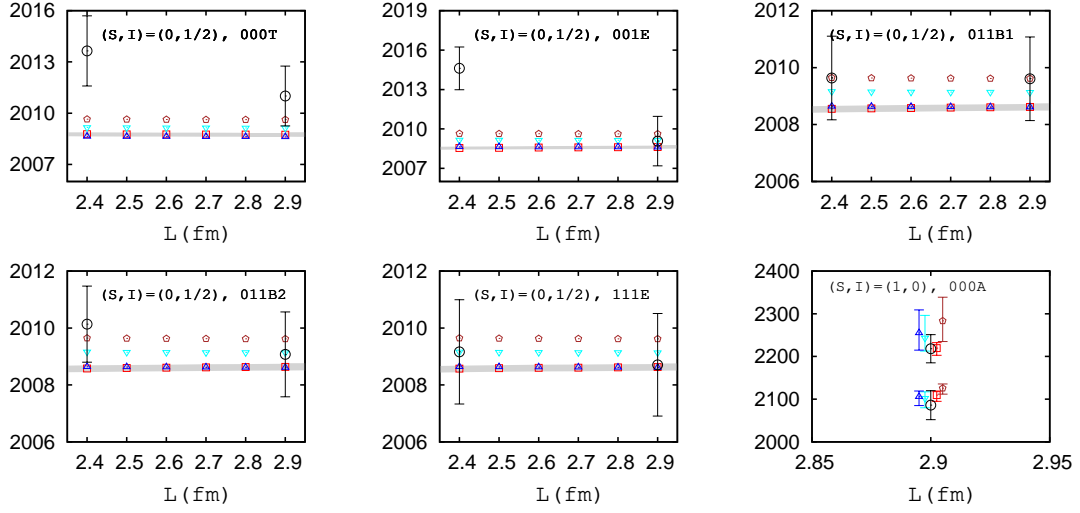} 
\caption{(Color online) Finite-volume energy levels of the $P$-wave $D\pi$ scattering with $(S,I)=(0,1/2)$ (the first five panels) and the energy levels of the elastic $S$-wave $DK$ scattering with $(S,I)=(1,0)$ (the last panel in the right bottom corner).  The lattice data for the $D\pi$ and $DK$ scattering are taken from Refs.~\cite{Moir:2016srx} and \cite{Lang:2014yfa}, respectively. For notations, see Fig.~\ref{fig.lev1}. }
   \label{fig.lev2}
\end{figure}

\begin{figure}[htbp]
\centering
\includegraphics[width=0.8\textwidth]{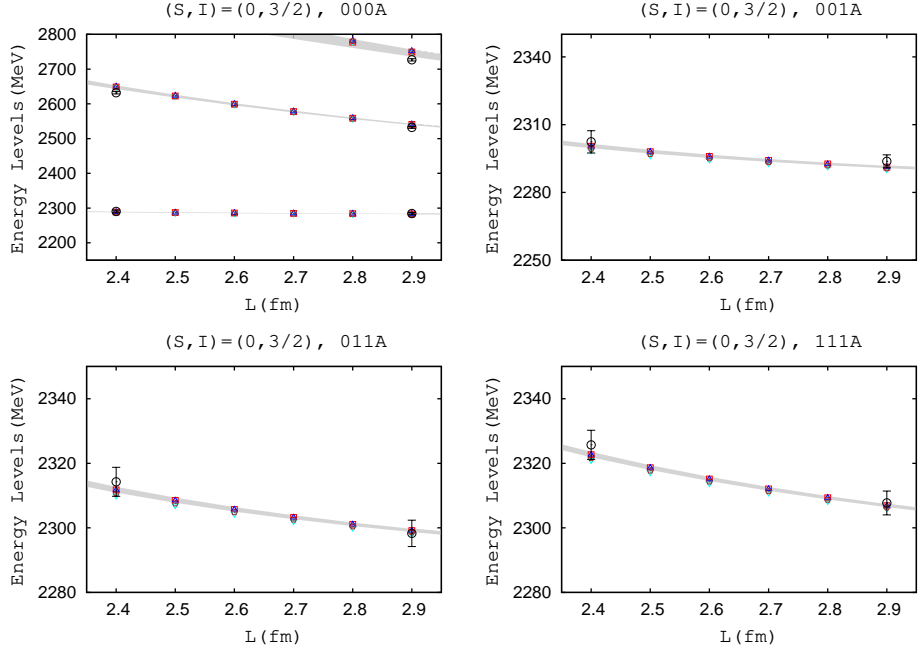} 
\caption{(Color online) Prediction of the finite-volume energy levels of the $S$-wave $D\pi$ scattering with $(S,I)=(0,3/2)$. The lattice data are extracted from Ref.~\cite{Moir:2016srx}. For notations, see Fig.~\ref{fig.lev1}. }
\label{fig.lev3}
\end{figure}

\begin{figure}[htbp]
\centering
\includegraphics[width=0.95\textwidth]{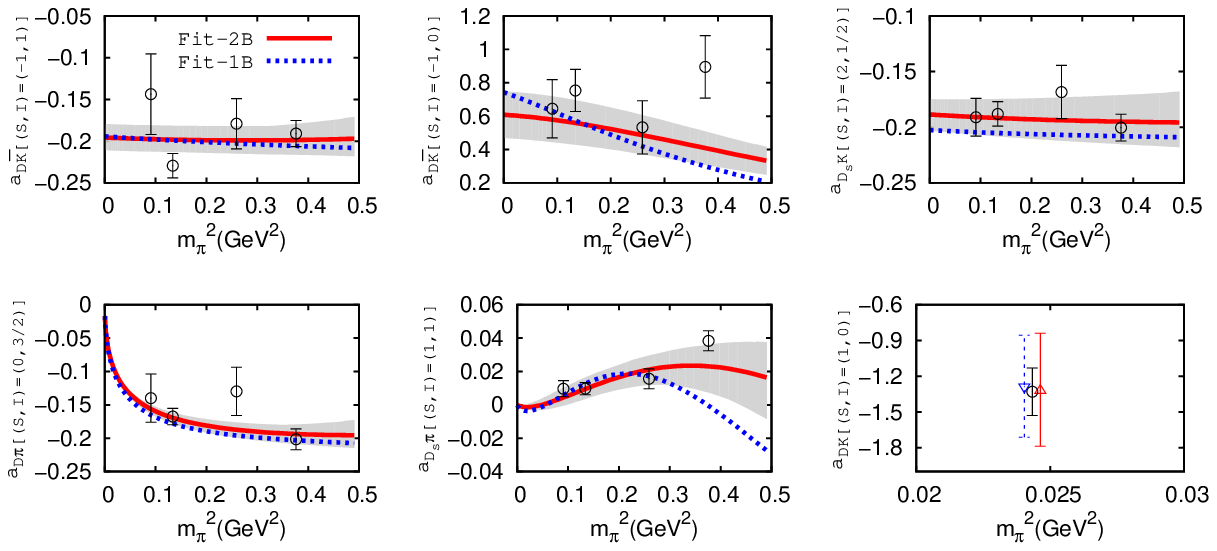} 
\caption{(Color online) Reproduction of various scattering lengths. The last panel in the right bottom corner denotes the results of the $S$-wave $DK$ scattering with $(S,I)=(1,0)$ and the lattice data are taken from Ref.~\cite{Lang:2014yfa}, where only the 2+1 flavor lattice simulation is considered. The lattice data in the other five panels are taken from Ref.~\cite{Liu:2012zya}. The red solid lines and the surrounding shaded areas correspond to the central results and the one-sigma error bands from Fit-2B. The blue dotted lines show the central-value results from Fit-1B.}\label{fig.sl}
\end{figure}

With all the parameters determined from the fits, we can reproduce the finite-volume energies of the scattering channels considered in this study. The reproduced energy levels as a function of the box size $L$ in various channels together with the lattice data are presented in Figs.~\ref{fig.lev1}, \ref{fig.lev2} and ~\ref{fig.lev3}. Fig.~\ref{fig.lev1} is for the $I=1/2$  coupled-channel scattering of $D\pi, D\eta$ and $D_s\bar{K}$. Fig.~\ref{fig.lev2} is for the $I=1/2$ $P$-wave $D\pi$ scattering and the $I=0$ $S$-wave $DK$ scattering. Fig.~\ref{fig.lev3} is for the $I=3/2$ $S$-wave $D\pi$ scattering, which is not included in the fits and is a prediction of our study. One can see that our theoretical formalism can well reproduce the lattice results. In these figures, we provide both the central values and one-sigma statistical error bands for Fit-2B. In order not to overload the figures, only the central-value results from Fit-1A, Fit-1B and Fit-2A are given. 

Similarly, we can also reproduce the scattering lengths given by the lattice calculations ~\cite{Liu:2012zya,Lang:2014yfa}. This is shown in Fig.~\ref{fig.sl}. Notice that only the data points with $m_\pi<600$~MeV from Ref.~\cite{Liu:2012zya} are included in the fits. We explicitly show the fit results from Fit-1B and Fit-2B, where $F_\pi$ is fixed at its physical value. Both the central values and one-sigma statistical error bands from Fit-2B are explicitly given. In order to not  overload the figures, we only show the central-value curves for Fit-1B. It is clear that our theoretical formalism can also well describe the various scattering lengths from the lattice calculations. 

Having determined all the unknown parameters and verified the reliability of our fits, we proceed to discuss the resonance structures in the scattering amplitudes in the next section.

\section{Phenomenological discussions in the infinite volume}\label{sect.pheno} 

\subsection{Scattering amplitudes and resonances at unphysical meson masses}

In this part, we study the infinite-volume amplitudes of the $D\pi, D\eta$ and $D_s\bar{K}$ coupled-channel scattering obtained at the unphysically large meson masses used in Ref.~\cite{Moir:2016srx}. The phase shifts ($\delta$) and inelasticity parameters ($\varepsilon$) are related to the $S$ matrix, which is given by the unitarized scattering amplitude $T$ in  Eq.~\eqref{eq.defut} through 
\begin{equation}
 S  = 1 + 2 i \sqrt{\rho(s)}\cdot T(s)\cdot \sqrt{\rho(s)}\,,  
\end{equation} 
with 
\begin{eqnarray}\label{eq.defrho}
 \rho(s)=\frac{\sigma(s)}{16\pi s}\,.
\end{eqnarray}
To be more specific, the phase shifts $\delta_{kk}$ and $\delta_{kl}$ and the inelasticity parameters $\varepsilon_{kk}$ and   $\varepsilon_{k l}$, with $k\neq l$, are related to the matrix elements $S_{kk}$ and $S_{k l}$ through   
\begin{align}\label{eq.defsmat} 
S_{ k k} = \varepsilon_{k k} {\rm e}^{2 i \delta_{ k k}}\,, \qquad 
S_{ k l} = i \varepsilon_{k l} {\rm e}^{ i \delta_{ k l}}\,.
\end{align}
For the inelasticity parameters $\varepsilon_{kk}$, one has $0\leq \varepsilon_{kk}\leq 1$.

\begin{figure}[t!]
\centering
\includegraphics[width=1.0\textwidth]{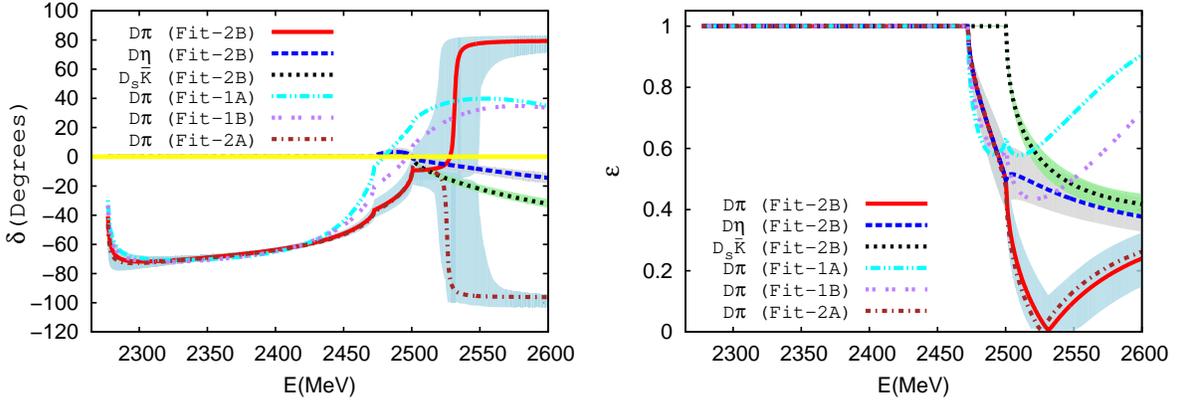} 
\caption{(Color online) Phase shifts ($\delta$) and inelasticities ($\varepsilon$) of the coupled-channel $D\pi, D\eta$ and $D_s\bar{K}$ $S$-wave scattering with $(S,I)=(0,1/2)$  obtained at $m_\pi=391$~MeV as in Ref.~\cite{Moir:2016srx}. The red solid, blue dashed and black dotted lines in the left panel denote the phase shifts obtained with the central values from Fit-2B  for the $D\pi, D\eta$ and $D_s\bar{K}$ channels, respectively. The surrounding shaded areas denote the uncertainties at the one-sigma level. The right panel shows the results for the inelasticities. In order to not overload the figures, we only give the central-value curves for the $D\pi$ channel from Fit-1A, Fit-1B and Fit-2A. }
   \label{fig.phasedpilatmass}
\end{figure}

The phase shifts and inelasticities of the $S$-wave coupled-channel $D\pi$, $D\eta$ and $D_s\bar{K}$ scattering with $(S,I)=(0,1/2)$ are given in the left and right panels in Fig.~\ref{fig.phasedpilatmass}, respectively. We show the representative results with both central values and the statistical uncertainties at  the one-sigma level from Fit-2B. Within uncertainties, we observe that there are two branches of the $D\pi$ phase shifts in the energy region around $E>2530$~MeV. The two branches of phase shifts in fact correspond to similar physical dynamics, since they differ by 
       180~degrees. Although the $D\pi$ phase shifts around 2530~MeV show large uncertainties, the inelasticities in the same energy region almost vanish, indicating that the underlying dynamics of the $S$ matrix in this region shows a unique feature. The central-value plots and uncertainties for the phase shifts and inelasticities of the $D\eta$ and $D_s\bar{K}$ channels from Fit-2B are also shown in Fig.~\ref{fig.phasedpilatmass}. In order to not overload the figure, we only give the central-value phase shifts and inelasticities for the $D\pi$ channel from Fit-1A, Fit-1B and Fit-2A. Roughly speaking, the $D\pi$ phase shifts below the first inelastic channel from different fits are quite compatible. In the region above 2450~MeV, the $D\pi$ phase shifts and inelasticities start to show different behaviors from different fits. The results from Fit-1A and Fit-1B, which include the same lattice simulation data up to the $D_s\bar{K}$ threshold, show somewhat similar behaviors.
       The results from Fit-2A and Fit-2B, which includes the lattice data above the $D_s\bar{K}$ threshold from Ref.~\cite{Moir:2016srx},  are clearly different from those from Fit-1A and Fit-1B in the energy region above 2450~MeV. The resulting plots from Fit-2A and Fit-2B, which include the same data sets, are compatible with each other within the uncertainties. The scattering amplitudes in the inelastic energy region are clearly affected by the different data included in the fits. In contrast, the amplitudes in the elastic energy region show quite consistent behaviors.

In order to study the resonance poles, we need to perform the analytical continuation of the scattering amplitudes to the complex energy plane. In our formalism, this can be easily achieved by modifying the $G(s)$ function in  Eq.~\eqref{eq.gfunc}. For each channel, one can define two Riemann sheets (RS's) for the $G(s)$ function. The formula in Eq.~\eqref{eq.gfunc} is the corresponding expression on the first RS. The expression on the second RS is given by~\cite{Oller:1997ti} 
\begin{eqnarray}\label{eq.defg2ndrs}
G(s)^{\rm DR}_{\rm II}(s) = G(s)^{\rm DR} + i \frac{\sigma(s)}{8\pi s}\,,
\end{eqnarray}
with $G(s)^{\rm DR}$ and $\sigma(s)$ defined in Eqs.~\eqref{eq.gfunc} and \eqref{eq.defsigma}, 
respectively. In this convention $\sigma(s)$ has to be calculated with $\text{Im}\,\sigma(s)>0$ in the complex $s$ plane. This implies that the signs of $\rm{Im}\,G(s)$ along the real $s$ axis above the threshold on the first and second RS's are opposite. For a $n$-channel problem, one can then define $2^n$ RS's. The first RS will be denoted as $(+,+,+,\cdots,+)$. The second, third and fourth RS's are labeled as $(-,+,+,\cdots,+)$, $(-,-,+,\cdots,+)$ and $(+,-,+,\cdots,+)$, respectively. The plus and minus signs correspond to the $G(s)$ function of this channel evaluated on the first and second RS's, respectively. Apart from the pole position $s_P$ itself, one can also calculate the residues $\gamma$ at $s_P$, which are given by 
\begin{eqnarray}
 T(s)= -\lim_{s\to s_P}\frac{\gamma\gamma^{\rm T}}{s-s_P}\,,
\end{eqnarray}
with $\gamma$ an $n$ row vector and its transpose $\gamma^{\rm T}=(\gamma_1,\gamma_2,\cdots,\gamma_n)$. The residues 
correspond to the coupling strengths of the resonance pole to the interacting channels and encode important information of the resonance. 

With the pole position and its residues, one can then further discuss the composition of the resonance. In Ref.~\cite{Guo:2015daa}, the calculation for the compositeness coefficient is generalized to the resonances by extending Weinberg's bound-state compositeness relation~\cite{Weinberg:1962hj}. Then one can interpret the values of the compositeness $X$ as the probabilities to find the two-body components inside the resonances and bound states. The prescription to calculate the compositeness coefficient $X_i$ contributed by the $i$th channel in Ref.~\cite{Guo:2015daa} is 
\begin{eqnarray}\label{eq.defx}
 X_i= \big|\gamma_i\big|^2 \bigg|\frac{d G(s)_i}{ds}\bigg|^2_{s=s_P}\,,
\end{eqnarray}
where the function $G(s)$ is given in Eq.~\eqref{eq.gfunc} or ~\eqref{eq.defg2ndrs}, depending on the location of the pole. The total compositeness $X$ is then given by $X=\sum_{i} X_i$, with the sum only spanning on the open channels for the resonance in question, which are the channels below the real part of the pole. We mention that the prescription in Eq.~\eqref{eq.defx} only applies to the canonical resonance, in the sense that the resonance pole should reside in the RS that can be directly accessed from the physical RS. For a near-threshold bound state, Eq.~\eqref{eq.defx} recovers the  Weinberg  compositeness~\cite{Weinberg:1962hj}.  We refer to Ref.~\cite{Guo:2015daa} for further details and also to  Ref.~\cite{Oller:2017alp}, where the general framework for the calculation of the compositeness for poles is developed.

In Table~\ref{tab.latpole}, we give both the resonance pole positions and their residues for the coupled-channel $D\pi, D\eta$ and $D_s\bar{K}$ $S$-wave scattering with $(S,I)=(0,1/2)$ obtained at the unphysically large meson masses used in  Ref.~\cite{Moir:2016srx}. The most robust conclusion from Table~\ref{tab.latpole} is that there is a bound state just below the $D\pi$ threshold. Within uncertainties the four fits lead to compatible results for the pole positions and also the corresponding residues. Our determinations of the bound state pole are close to the value in Ref.~\cite{Moir:2016srx}, which is $2275.9\pm0.9$~MeV. Furthermore, the results from Fit-2A and Fit-2B, which include exactly the same lattice data of the $D\pi$ coupled-channel scattering as Ref.~\cite{Moir:2016srx}, are perfectly compatible with the value given in the former reference. This presents a nice crosscheck with our chiral amplitudes assisted finite-volume study, comparing with the $K$-matrix assisted L\"uscher formula in Ref.~\cite{Moir:2016srx}. While in Ref.~\cite{Albaladejo:2016lbb}, the bound state pole is predicted to be $2264_{-14}^{+8}$~MeV, with much larger error bars than ours and those in Ref.~\cite{Moir:2016srx}. The possible reason is that the values of the chiral LECs and the subtraction constant used in Ref.~\cite{Albaladejo:2016lbb} are taken from Ref.~\cite{Liu:2012zya}, which were determined by only including the elastic scattering lengths of the latter reference. Regarding the coupling strengths of the bound state, our study shows that this pole is more strongly coupled to the $D_s\bar{K}$ channel than to the $D\eta$ one, which is also the case of Ref.~\cite{Albaladejo:2016lbb} but differs from the results of Ref.~\cite{Moir:2016srx}. By applying Eq.~\eqref{eq.defx}, the compositeness coefficients of the bound state contributed by $D\pi, D\eta$ and $D_s\bar{K}$ are $0.91_{-0.02}^{+0.03}$, $0.01_{-0.00}^{+0.00}$ and $0.04_{-0.02}^{+0.02}$, respectively. Therefore we quantitatively verify that the $D\pi$ component overwhelmingly dominates the bound state in the $S$-wave $(S,I)=(0,1/2)$ channel at $m_\pi=391$~MeV.

The other robust pole is the one appearing between $2.4$ and $2.5$~GeV. All the four fits lead to the resonance pole on the third RS  with a mass around 2450~MeV and a half width around 130~MeV. For Fit-1A and Fit-1B, we also find shadow poles on the second RS, with the mass around 2490~MeV and a half width around 35~MeV. For Fit-2A and Fit-2B, not all of the parameter configurations within one-sigma uncertainties can generate a shadow resonance pole on the second RS around 2500~MeV. Since the poles on the second RS are mostly off the resonant ranges, we do not explicitly show their positions and residues in Table~\ref{tab.latpole}.  
Our results for the resonance poles on the third RS and their residues are consistent with the determinations in  Ref.~\cite{Albaladejo:2016lbb}. However the resonance poles, either on the second or the third RS's, are not reported in Ref.~\cite{Moir:2016srx}. Apart from the poles shown in Table~\ref{tab.latpole}, we find that there are also other heavier  poles in the region around or above 2600~MeV. The heavier poles appear in different RS's depending on the different fits and the uncertainties of their masses are usually large. Since these heavier resonances are much less constrained in our fits and show large model dependences, we refrain from discussing further about their properties. 

According to Ref.~\cite{Guo:2015daa}, the prescription in Eq.~\eqref{eq.defx} can be applied to the third-sheet poles from Fit-2A and Fit-2B in Table~\ref{tab.latpole}, since they are above the $D\eta$ threshold. The resulting compositenesses of the pole from Fit-2A are $0.17_{-0.02}^{+0.02}$ ($D\pi$) and $0.29_{-0.04}^{+0.04}$ ($D\eta$). The results for the pole from Fit-2B are $0.20_{-0.03}^{+0.03}$ ($D\pi$) and $0.29_{-0.05}^{+0.06}$ ($D\eta$). Therefore, we can conclude that the resonance pole around $2.4$~GeV obtained at unphysically large meson masses contain other important components apart from the $D\pi$ and $D\eta$.

For the $P$-wave $D\pi$ scattering, a bound state pole with the mass in the range $2008.2\sim2009.8$~MeV is found by combining the results from the four fits. Our determination is in good agreement with the value $2009\pm2$~MeV in  Ref.~\cite{Moir:2016srx}. 

Having shown the scattering amplitudes and the resonance structures at the unphysical meson masses, we proceed the study for the physical meson masses in the following section.

\begin{table}[htbp]
\centering
\begin{footnotesize}
\begin{tabular}{c c c c c c c }
\hline\hline
    Fit &  RS & M~(MeV) & $\Gamma$/2~(MeV) & $|\gamma_{1}|$~(GeV) & $|\gamma_{2}/\gamma_{1}|$ & $|\gamma_{3}/\gamma_{1}|$
\\ \hline\hline
  Fit-1A & I & $2275.0_{-1.0}^{+0.9}$ & $0$ &$4.8_{-1.0}^{+0.6}$  & $0.31_{-0.10}^{+0.08}$  &$0.92_{-0.09}^{+0.08}$  
\\  
Fit-1A & III & $2430.5_{-52.4}^{+46.1}$ & $119.7_{-33.7}^{+27.1}$ &$7.8_{-0.9}^{+0.7}$  & $0.92_{-0.08}^{+0.11}$  &$1.84_{-0.21}^{+0.29}$  
\\ \hline
Fit-1B & I & $2275.4_{-1.1}^{+0.7}$ & $0$ &$4.5_{-1.0}^{+0.8}$  & $0.33_{-0.09}^{+0.08}$  &$0.93_{-0.08}^{+0.05}$    
\\  
Fit-1B & III & $2432.2_{-48.6}^{+59.0}$ & $157.8_{-30.1}^{+30.4}$ &$8.2_{-0.9}^{+1.2}$  & $0.79_{-0.06}^{+0.08}$  &$1.85_{-0.24}^{+0.21}$  
\\ \hline
Fit-2A & I & $2276.1_{-0.6}^{+0.4}$ & $0$ &$3.6_{-1.9}^{+0.9}$  & $0.09_{-0.04}^{+0.04}$  &$0.70_{-0.03}^{+0.03}$   
\\ 
Fit-2A & III     & $2490.9_{-21.9}^{+24.8}$ & $104.2_{-17.1}^{+23.6}$ &$7.1_{-0.6}^{+0.6}$  & $1.00_{-0.09}^{+0.10}$  &$1.82_{-0.15}^{+0.16}$   
\\ \hline
Fit-2B & I & $2275.8_{-0.8}^{+0.5}$ & $0$ &$3.9_{-1.4}^{+0.8}$  & $0.14_{-0.05}^{+0.04}$  &$0.70_{-0.04}^{+0.04}$   
\\ 
Fit-2B & III     & $2486.7_{-28.5}^{+29.9}$ & $126.6_{-21.5}^{+29.0}$ &$7.7_{-0.6}^{+0.6}$  & $0.95_{-0.11}^{+0.11}$  &$1.79_{-0.14}^{+0.15}$   
\\
\hline\hline
\end{tabular}
\end{footnotesize}
\caption{\label{tab.latpole} Poles and their residues obtained at unphysical meson masses ($m_\pi=391$~MeV) used in the lattice simulation~\cite{Moir:2016srx} from the $S$-wave coupled $D\pi, D\eta$ and $D_s\bar{K}$ scattering amplitudes with $(S,I)=(0,1/2)$. The indices of the residues $\gamma_{i=1,2,3}$ correspond to the channels $D\pi, D\eta$ and $D_s\bar{K}$ in order. The thresholds of the $D\pi, D\eta$ and $D_s\bar{K}$ channels are $2276.5, 2472.4$ and $2500.5$~MeV, respectively. For the definition of different RS's, see the text for details.}
\end{table} 

\subsection{Chiral extrapolation to the physical meson masses}

\begin{figure}[htbp]
\centering
\includegraphics[width=1.0\textwidth]{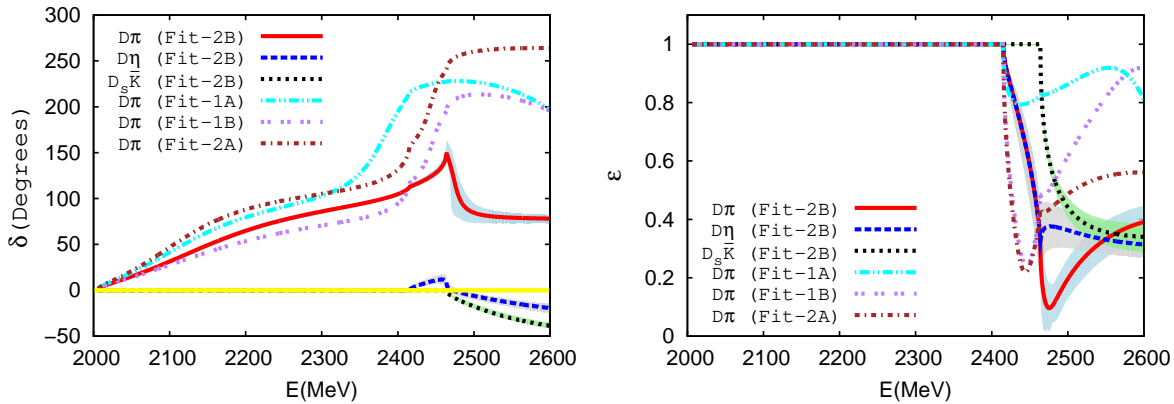} 
\caption{(Color online) Phase shifts ($\delta$) and inelasticities ($\varepsilon$) of the coupled-channel $D\pi, D\eta$ and $D_s\bar{K}$ $S$-wave scattering with $(S,I)=(0,1/2)$  obtained at physical meson masses. For notaion, see Fig.~\ref{fig.phasedpilatmass}. }
   \label{fig.phasedpiphymass}
\end{figure}

\begin{figure}[htbp]
\centering
\includegraphics[width=1.0\textwidth]{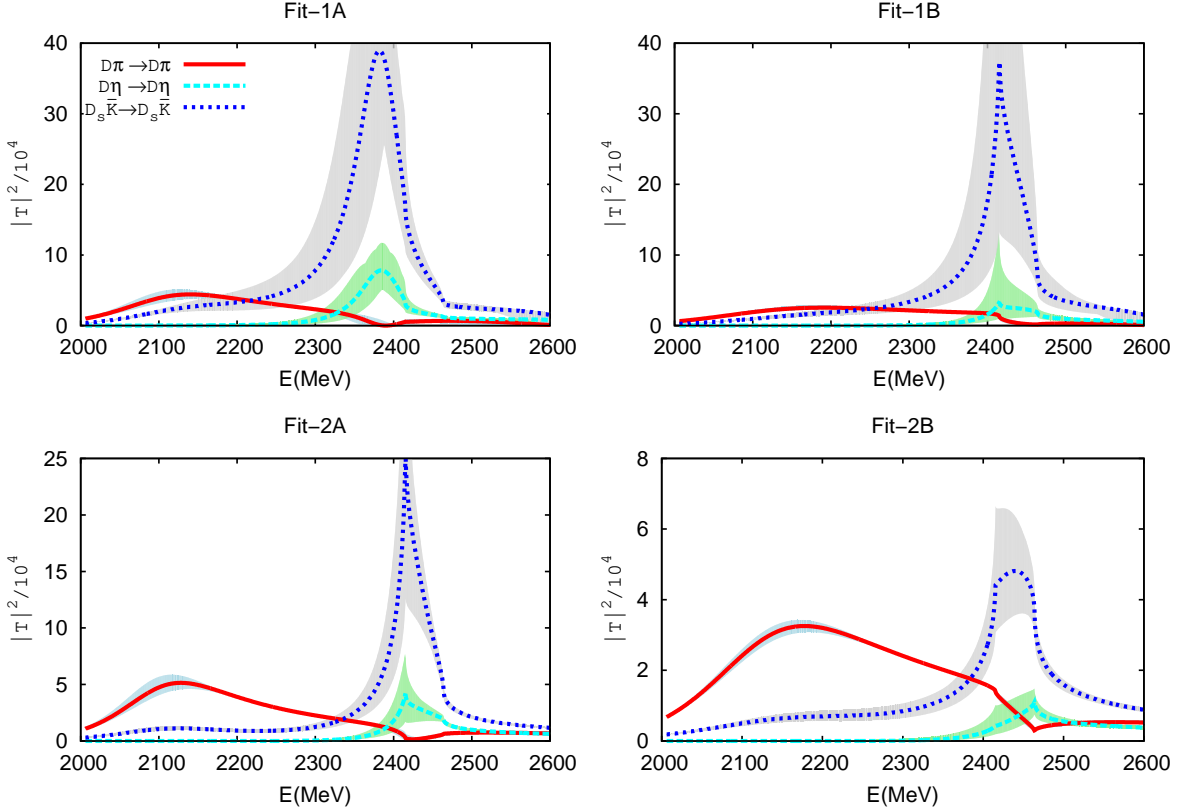} 
\caption{(Color online) Magnitudes squared of the scattering amplitudes of the $S$-wave $D\pi, D\eta$ and $D_s\bar{K}$ scattering with $(S,I)=(0,1/2)$ obtained at physical meson masses from the four types of fits. The red solid, cyan dashed and blue dotted lines denote the $D\pi\to D\pi$, $D\eta\to D\eta$ and $D_s\bar{K}\to D_s\bar{K}$ amplitudes, respectively. The surrounding shaded areas correspond to the uncertainties at the one-sigma level.  }
   \label{fig.tdkphymass}
\end{figure}

By assuming that the free parameters in Table~\ref{tab.fitlecs} are independent on the light-flavor meson masses, it is straightforward to perform the chiral extrapolation to the physical meson masses in our study. The phase shifts and inelasticities of the $S$-wave coupled-channel $D\pi$, $D\eta$ and $D_s\bar{K}$ scattering obtained at physical meson masses are given in Fig.~\ref{fig.phasedpiphymass}. The central-value plots and the statistical uncertainties at  the one-sigma level from Fit-2B are shown. As in Fig.~\ref{fig.phasedpilatmass}, we give the central-value curves for Fit-1A, Fit-1B and Fit-2A in Fig.~\ref{fig.phasedpiphymass}. In order to clearly demonstrate the resonance structures, the magnitudes of the scattering $T$ matrices are provided in Fig.~\ref{fig.tdkphymass}. One can clearly see the discrepancies from different fits.

Another subtlety issue on the chiral extrapolation is about the pion  mass dependences of the parameters in Table~\ref{tab.fitlecs}. 
Note that the strange-quark mass is basically kept fixed to its physical value in the lattice QCD here considered.
 Clearly the chiral LECs by definition are independent of the pion mass.
The subtraction constant $a$, on the contrary, could possibly vary with different pion masses,
although many previous works simply assume the constant behavior of $a$ when performing the chiral  extrapolation~\cite{Yao:2015qia,Guo:2015dha,Du:2017ttu,Altenbuchinger:2013gaa,Altenbuchinger:2013vwa,Wang:2012bu,Liu:2012zya, Albaladejo:2016lbb}.
One possible way to estimate the pion mass dependences is by comparison of 
the function $G^{DR}(s)$ in Eq.~\eqref{eq.gfunc}  to the three-momentum-cutoff version of $G(s)$ \cite{Oller:1998hw}.

First, the on-shell three-momentum is denoted by $q(s)$ with 
\begin{align}
  \label{181006.1}
  q(s)&=\frac{\lambda(s,m_1^2,m_2^2)}{2\sqrt{s}}~.
\end{align}
We introduce the function $\delta G(s)$ by  rewriting $G^{DR}(s)$ in Eq.~\eqref{eq.gfunc} as
\begin{align}
  \label{181006.2}
 G^{DR}(s)&=\frac{a}{16\pi^2}+\delta G(s)~.
\end{align}
Next, we denote by $G^C(s)$ the function that results by evaluating the 
divergent integral of Eq.~\eqref{eq.defg} with a three-momentum cutoff $q_{\rm max}$. 
An algebraic expression for $G^C(s)$ can be found in Ref.~\cite{Oller:1998hw}

Now, let us consider the possible pion mass dependence of the subtraction constants. One can work out explicitly the (nonrelativistic) limit of the functions
$G^{DR}(s)$, Eq.~\eqref{eq.gfunc}, and $G^C(s)$ \cite{Oller:1998hw} for $s\to (m_1+m_2)^2$, which 
implies $|q(s)|\ll m_1,\,m_2, \,q_{\rm max}$.  In this limit these functions are simply 
a constant plus $-i q(s)/8\pi (m_1+m_2)$ plus quadratic and higher-order terms in three-momentum.
Therefore, we can write 
\begin{align}
  \label{181006.3}
 G^{DR}(s)- G^{C}(s)&=\frac{\alpha}{16\pi^2}+{\cal O}(q^2)~,
\end{align}
where $\alpha$ is a constant. 
For the case $q(s)=0$, the following expression for the subtraction constant $a$ is obtained
\begin{align}
  \label{181006.4}
  a(\mu)&=\alpha+G^C((m_1+m_2)^2)-\delta G((m_1+m_2)^2)\nn\\
  &=\alpha-2\frac{1}{m_1+m_2}\left[
  m_1 \log\left(1+\sqrt{1+\frac{m_1^2}{q_{\rm max}^2}}\right)
  +m_2 \log\left(1+\sqrt{1+\frac{m_2^2}{q_{\rm max}^2}}\right)\right]
  +\log\frac{\mu^2}{q_{\rm max}^2}~.
\end{align}
The constant $\alpha$ does not depend on $\mu$, since $a(\mu)-\log\mu^2$ is $\mu$ independent, cf. Eq.~\eqref{eq.gfunc}. Eq.~\eqref{181006.4}  reflects the splitting in the mechanisms underlying the
contributions to the subtraction constant $a$.  On the one hand, the last two contributions in this equation stem from
rescattering effects (by taking $q_{\max}$ around 1~GeV), associated with the unitarity cut.
On the other hand, the former contribution ($\alpha$)  is associated 
with properties of contact terms (short-range physics). 
 The crossed-channel contributions involving the
explicit degrees of freedom in the effective field theory will be accounted for order by order
in $\mV(s)$ \eqref{eq.defut}.
Thus, the variation  of $\alpha$ with the masses is  at least quadratic in the pion mass.
There is some remnant cut-off dependence in the splitting of Eq.~\eqref{181006.4} that could be ascertained 
by varying $q_{\rm max}$ around $q_{\rm max}\simeq 1$~GeV,
the typical scale for hadronic  rescattering.

The fact that the leading correction to Eq.~\eqref{181006.4} is linear in $m_2$
implies a linear change in the pion mass for the $D\pi$ subtraction constants, as $m_\pi/m_D$.
 Differently, for the other channels involving a $K$ or an $\eta$ the
change will be just quadratic in $m_\pi$ and, therefore, much less important.
 The linearized version of Eq.~\eqref{181006.4} with respect to the smaller mass $m_2$
for two sets of values of the masses $m_1$ and $m_2$ ($a'$ and $a$ for $m_i'$ and $m_i$, in order) gives 
\begin{align}
  \label{181006.6}
  a'-a&=2\frac{m_2'-m_2}{m_1} \log\frac{1+\sqrt{1+m_1^2/q_{\rm max}^2}}{2}+{\cal O}(m_2^2)~.
\end{align}

The subtraction constant $a$ is said to have its natural value   \cite{Oller:2000fj} when 
the constant $\alpha$ has an absolute value much smaller than 1 for $q_{\rm max}\simeq 1~$GeV.
One then has \cite{Oller:2000fj}
\begin{align}
  \label{181006.5}
  a(q_{\rm max})&=-2\log\left(1+\sqrt{1+\frac{m_1^2}{q_{\rm max}^2}}\right)+\ldots \simeq -2.3~,
\end{align}
where the ellipses indicates higher order terms in the nonrelativistic expansion and in $m_2/m_1$ with $m_2\ll m_1$, as it follows directly from Eq.~\eqref{181006.4}.

\begin{figure}[htbp]
\centering
\includegraphics[width=1.0\textwidth]{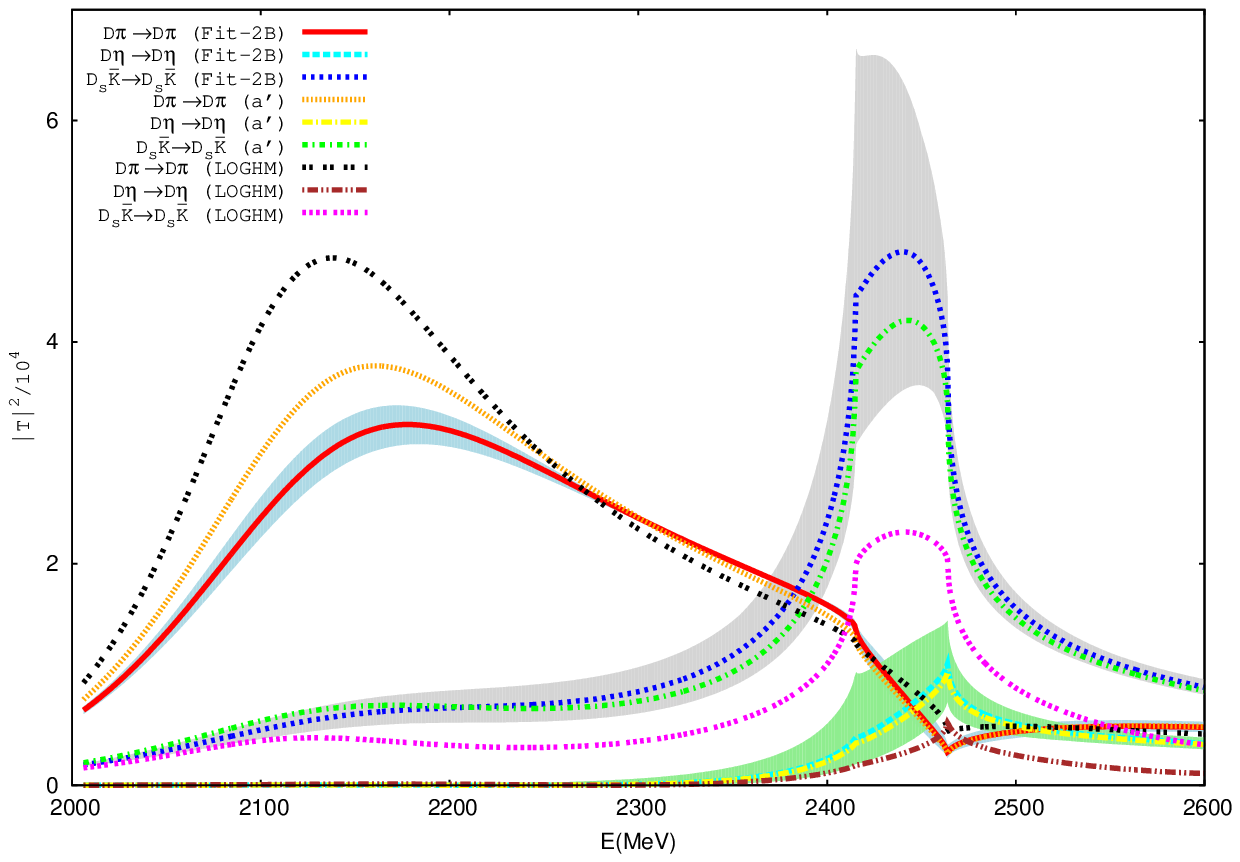} 
\caption{(Color online) Comparison of the magnitudes squared of the $S$-wave amplitudes of the $D\pi, D\eta$ and $D_s\bar{K}$ scattering with $(S,I)=(0,1/2)$ at physical meson masses using the parameters from Ref.~\cite{Liu:2012zya} and Fit-2B w/o including the pion mass correction to the subtraction constant $a_{D\pi}^{0,1/2}$. The results from Fit-2B without and with introducing the pion mass correction to $a_{D\pi}^{0,1/2}$ are labeled as Fit-2B and $a'$, respectively. See the text for details. The results using the parameters from Ref.~\cite{Liu:2012zya} are labeled as LOGHM.  }
   \label{fig.tdkphymasscompare}
\end{figure}

We take the Fit-2B as a concrete example to check the shifts of $a$ by varying pion masses.
When Eq.~\eqref{181006.6} is applied to a $D\pi$ subtraction constant from a pion mass of 391~MeV to its physical value of $138$~MeV
we have a variation in the subtraction constant of $-0.12$. 
Compared to the value reported in Table~\ref{tab.fitlecs} for $a_{D\pi}^{0,1/2}$, we obtain a mild effect of around a 10\%.
For the subtraction constants in the $D\eta$ and $D_s\bar{K}$ channels, their values are kept fixed.
In Fig.~\ref{fig.tdkphymasscompare}, we explicitly show the results by taking the extrapolated value $a_{D\pi}^{0,1/2}=-1.64$ obtained from Eq.~\eqref{181006.6}, together with the figures by assuming pion mass independence of the subtraction constants from Fit-2B and also the results from Ref.~\cite{Liu:2012zya}. The three different sets of plots in Fig.~\ref{fig.tdkphymasscompare} reveal qualitatively similar resonant behaviors, although the heights of the peaks around the resonances at $2.1$~GeV and $2.45$~GeV are different.
 Comparing the curves from Fit-1A, Fit-1B and Fit-2A in Fig.~\ref{fig.tdkphymass} with those in Fig.~\ref{fig.tdkphymasscompare},
 we observe that the discrepancies among the  
 three different types of plots in Fig.~\ref{fig.tdkphymasscompare}, which include the additional uncertainties of the chiral extrapolation
 and the fitting results from the previous work~\cite{Liu:2012zya}, are clearly smaller than the differences of the four types
 of fits in the present work.
 Therefore in the following discussions, we shall  concentrate on the results from the four types of fits in Table~\ref{tab.fitlecs}
 without introducing the pion mass corrections to the subtraction constants.

\begin{figure}[t!]
\centering
\includegraphics[width=1.0\textwidth]{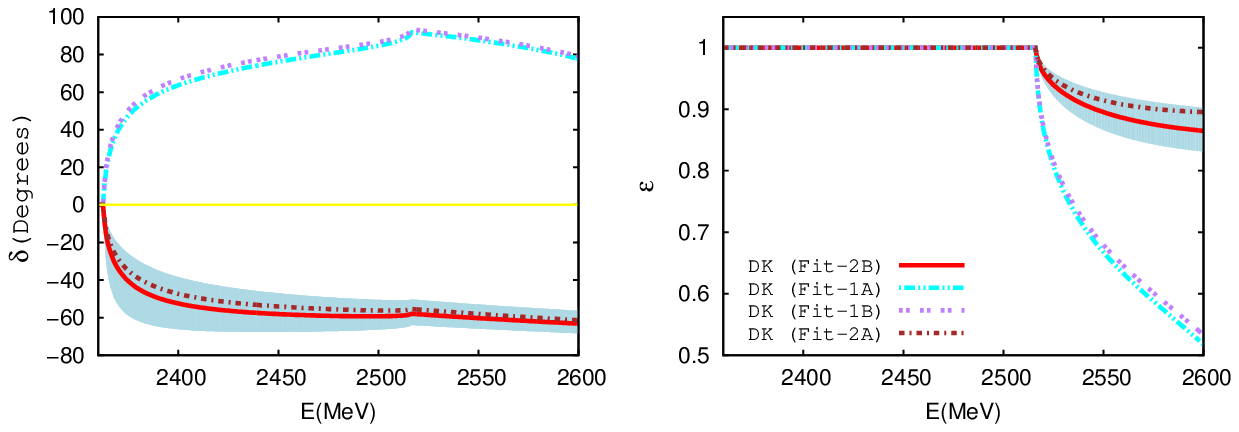} 
\caption{(Color online) Phase shifts ($\delta$) and inelasticities ($\varepsilon$) of the $S$-wave $DK$ scattering with $(S,I)=(1,0)$ obtained at physical meson masses. The red solid lines and the surrounding areas correspond to the central values and uncertainties at the one-sigma level obtained from Fit-2B, respectively. Similar as  Figs.~\ref{fig.phasedpilatmass} and \ref{fig.phasedpiphymass}, only the central-value results are shown for Fit-1A and Fit-1B. }
   \label{fig.phasedkphymass}
\end{figure}

In Fig.~\ref{fig.phasedkphymass}, we show the phase shifts and inelasticities of the $DK$ channel obtained at physical meson masses. In this case, the central-value results from Fit-1A and Fit-1B are almost identical, which lead to a sharp rise of the phase shifts near the $DK$ threshold. The central-value behavior from Fit-1A and Fit-1B indicates a virtual pole near the $DK$ threshold. In contrast, the phase shifts from Fit-2A and Fit-2B fall rapidly, which implies a bound state pole near the threshold.\footnote{This statement is clear if one considers
  an effective range expansion including only the scattering length $a$, so that $t(q)=1/(1/a-iq)$, with $q$
the CM three-momentum. The case $a<\!(>)0$ corresponds to a bound (virtual) state.}

The resonance poles and the corresponding residues from the $S$-wave coupled-channel $D\pi$, $D\eta$ and $D_s\bar{K}$ scattering with $(S,I)=(0,1/2)$ obtained at physical meson masses are collected in Table~\ref{tab.phypoledpi}. The first lesson we learn is that all the four fits give robust resonance poles on the second RS with the mass around 2100~MeV and the half width lying between 100 and 200~MeV. Furthermore, for all the four fits we also find their shadow poles on the third RS, whose masses and widths are quite close to the values on the second RS. Heavier resonance poles lying around $2.4$~GeV are found as well. In Table~\ref{tab.phypoledpi}, the relevant poles that are mostly responsible for the resonant behaviors on the physical sheet are given. For Fit-1A, the relevant pole is located on the second RS, and its shadow pole lies on the third RS. While for the other three fits, the relevant poles lie on the third RS and the corresponding shadow poles are found on the second RS. The masses and half widths of the shadow poles around $2.4$~GeV on other RS's are somewhat different from the relevant poles. For example, the shadow pole from Fit-1A lies on the third RS, with the pole position $(2291.3_{-41.6}^{+49.1} - i\,54.9_{-20.4}^{+22.2})$~MeV. Both the shadow poles from Fit-1B and Fit-2A are found on the second RS, with the positions of $(2445.2_{-18.9}^{+23.7} - i\,12.3_{-7.8}^{+7.3})$~MeV and $(2443.1_{-12.1}^{+13.7} - i\,12.0_{-6.5}^{+8.5})$~MeV, respectively. For the case of Fit-2B, only the relevant poles on the third sheet are found and we do not see the nearby shadow poles on other RS's. In the energy region around $2.4$~GeV we only observe bumps, instead of poles, on the second RS for Fit-2B. The resonance contents from Fit-2B resemble the results in Ref.~\cite{Albaladejo:2016lbb}. 
We stress that the poles and their residues in Table~\ref{tab.phypoledpi} are only slightly affected by the pion mass dependences of the subtraction constants. We also take Fit-2B to demonstrate this point.  Taking the chiral extrapolated value $a_{D\pi}^{0,1/2}=-1.64$ as explained previously, the pole around $2.1$~GeV on the second RS is found at $(2112.5-i 127.0)$~MeV, with the residues $|\gamma_1|=9.9$~GeV, $|\gamma_2/\gamma_1|=0.08,\,|\gamma_3/\gamma_1|=0.58$. The pole around $2.4$~GeV on the third RS is  at $(2475.7-i 108.9)$~MeV, with the residues $|\gamma_1|=6.7$~GeV, $|\gamma_2/\gamma_1|=1.15,\, |\gamma_3/\gamma_1|=2.07$. These results are consistent with the values from Fit-2B that assumes the pion mass independence of the subtraction constants given in Table~\ref{tab.phypoledpi}.

Next we discuss the compositeness for the resonance poles obtained at physical meson masses. According to Ref.~\cite{Guo:2015daa}, the prescription in Eq.~\eqref{eq.defx} can be applied to the poles around $2.1$~GeV on the second RS in Table~\ref{tab.phypoledpi}. The compositeness coefficients contributed by the $D\pi$ channel in Fit-1A, Fit-1B, Fit-2A and Fit-2B are $0.43_{-0.02}^{+0.02}$, $0.47_{-0.01}^{+0.03}$, $0.42_{-0.01}^{+0.01}$ and $0.47_{-0.01}^{+0.01}$, respectively. This implies that both the $D\pi$ component and other degrees of freedom are important for the broad scalar resonance around $2.1$~GeV in the $(S,I)=(0,1/2)$ channel. Eq.~\eqref{eq.defx} is also valid for the second sheet pole around $2.4$~GeV from Fit-1A, which gives the $D\pi$ compositeness $0.06_{-0.02}^{+0.02}$, indicating that the role of the $D\pi$ in the scalar resonance pole around $2.4$~GeV is marginal. Although, rigorously speaking, the prescription in Eq.~\eqref{eq.defx} can not be applied to other poles in Table~\ref{tab.phypoledpi}, the poles on the third RS from Fit-1B, Fit-2A and Fit-2B are not far away from the  region of validity according to Ref.~\cite{Guo:2015daa}. As a rough estimate, we also use Eq.~\eqref{eq.defx} to calculate the compositeness coefficients for those poles. The compositenesses for the pole from Fit-1B are $0.10_{-0.02}^{+0.02}$ ($D\pi$) and $0.38_{-0.09}^{+0.11}$ ($D\eta$). The corresponding values for the pole from Fit-2A are $0.06_{-0.01}^{+0.01}$ ($D\pi$) and $0.35_{-0.05}^{+0.06}$ ($D\eta$). For the pole from Fit-2B, the compositeness coefficients are $0.11_{-0.02}^{+0.02}$ ($D\pi$) and $0.31_{-0.04}^{+0.04}$ ($D\eta$). This tells us that the other degrees of freedom beyond the $D\pi$ and $D\eta$ components play important roles in the scalar charmed resonance pole around $2.4$~GeV.

\begin{table}[htbp]
\centering
\begin{footnotesize}
\begin{tabular}{c c c c c c c }
\hline\hline
Fit &  RS & M & $\Gamma$/2~(MeV)  & $|\gamma_1|$~(GeV)  & $|\gamma_2/\gamma_1|$ & $|\gamma_3/\gamma_1|$    \\
\hline\hline
Fit-1A & II & $2097.7_{-6.1}^{+6.8}$ & $112.2_{-14.2}^{+16.5}$ &$9.6_{-0.3}^{+0.3}$  & $0.10_{-0.04}^{+0.05}$  &$0.78_{-0.08}^{+0.08}$  
\\ 
Fit-1A & II & $2384.4_{-23.6}^{+26.4}$ & $36.0_{-10.0}^{+9.9}$ &$4.8_{-0.6}^{+0.5}$  & $1.51_{-0.16}^{+0.15}$  &$2.09_{-0.18}^{+0.18}$  
\\ \hline
Fit-1B & II & $2106.4_{-5.0}^{+5.1}$ & $170.6_{-13.0}^{+12.5}$ &$10.1_{-0.2}^{+0.3}$  & $0.11_{-0.07}^{+0.07}$  &$0.79_{-0.07}^{+0.07}$   
\\
Fit-1B & III & $2409.0_{-24.5}^{+22.7}$ & $78.6_{-15.2}^{+20.5}$ &$6.1_{-0.6}^{+0.7}$  & $1.22_{-0.19}^{+0.19}$  &$2.72_{-0.49}^{+0.48}$   
\\ \hline
Fit-2A & II & $2095.7_{-6.8}^{+5.2}$ & $97.1_{-10.7}^{+10.3}$ &$9.4_{-0.2}^{+0.2}$  & $0.10_{-0.02}^{+0.02}$  &$0.63_{-0.03}^{+0.03}$   
\\ 
Fit-2A & III & $2401.3_{-19.6}^{+20.4}$ & $55.0_{-10.8}^{+14.5}$ &$5.1_{-0.5}^{+0.5}$  & $1.31_{-0.15}^{+0.19}$  &$2.50_{-0.28}^{+0.31}$   
\\ \hline
Fit-2B & II & $2117.7_{-3.4}^{+3.8}$ & $145.0_{-6.8}^{+8.0}$ &$10.2_{-0.1}^{+0.2}$  & $0.09_{-0.03}^{+0.03}$  &$0.58_{-0.03}^{+0.04}$   
\\
Fit-2B & III & $2470.5_{-24.9}^{+25.1}$ & $104.1_{-12.5}^{+16.0}$ &$6.7_{-0.6}^{+0.7}$  & $1.14_{-0.12}^{+0.12}$  &$2.06_{-0.16}^{+0.16}$  
\\
\hline\hline
\end{tabular}
\end{footnotesize}
\caption{\label{tab.phypoledpi}  Poles and their residues obtained at physical meson masses from the $S$-wave coupled-channel $D\pi, D\eta$ and $D_s\bar{K}$ scattering with $(S,I)=(0,1/2)$. The physical thresholds of the $D\pi, D\eta$ and $D_s\bar{K}$ channels are $2005.3, 2415.1$ and $2463.9$~MeV, respectively. For the definitions of different RS's and notations, see the text and Table~\ref{tab.latpole} for details. }
\end{table}

For the $\ds$ in the $S$-wave coupled-channel $DK$ and $D_s\eta$ scattering, its poles and residues obtained at physical meson masses are summarized in Table~\ref{tab.phypoledk}. First we stress that each of the parameter configurations from all of the fits only gives one pole  for the $\ds$, either a bound state or a virtual state. For the parameters from Fit-2A and Fit-2B within one-sigma uncertainties, all the parameter configurations only give the bound state poles on the first RS. While for  Fit-1A and Fit-1B, within one-sigma uncertainties parts of the parameter configurations give the bound state poles on the first RS and others give the virtual poles on the second RS. E.g., with the central values of the parameters from Fit-1A and Fit-1B we only obtain the virtual poles. This tells us that the interactions are strong enough to produce a prominent enhancement below the $DK$ threshold. However at the present stage we can not definitely conclude that the enhancement is caused by a bound  or a virtual state. 
The findings of the bound and virtual states are consistent with the behaviors of the phase shifts shown in Fig.~\ref{fig.phasedkphymass}. 

For the bound state poles, we can use  Eq.~\eqref{eq.defx} to calculate the compositeness coefficients for the $\ds$. 
The compositeness coefficients contributed by the $DK$ and $D_s\eta$ for the Fit-2A case are $0.72_{-0.13}^{+0.14}$ and $0.16_{-0.07}^{+0.04}$, respectively. The corresponding values from Fit-2B are $0.77_{-0.13}^{+0.11}$ and $0.11_{-0.04}^{+0.03}$, which are compatible with those found in Ref.~\cite{Guo:2015daa}. The robust conclusion from these numbers is that the $DK$ component is the dominant one inside the $\ds$.

Before ending the phenomenological discussion, we give the predictions for the $S$-wave scattering lengths of various channels at the physical meson masses in Table~\ref{tab.sl}. Only the central values of the $D\bar{K}$ scattering lengths with $(S,I)=(-1,0)$ from Fit-1A and Fit-2A and the $DK$ scattering lengths with $(S,I)=(1,0)$ from Fit-1A and Fit-1B, which are marked with asterisks, are given. For other entries in Table~\ref{tab.sl}, we provide the values with statistical uncertainties. This is because within the one-sigma fitted parameter configurations the scattering lengths for these four channels vary from huge negative values to huge positive values. The reason behind is that for these channels parts of the parameter configurations could lead to bound state poles near threshold, which correspond to large negative scattering lengths,  and others could give virtual poles near threshold, which correspond to large positive scattering lengths. These findings are consistent with the pole contents discussed in Table~\ref{tab.phypoledk} for the $\ds$. We also verify that similar situations happen for the $S$-wave $D\bar{K}$ scattering with $(S,I)=(-1,0)$. The results from Fit-2B in Table~\ref{tab.sl}, which gives the closest values of the parameters to Refs.~\cite{Guo:2015dha,Liu:2012zya}, are qualitatively compatible with the numbers of the former references within uncertainties. 

\begin{table}[htbp]
\centering
\begin{footnotesize}
\begin{tabular}{c c c c c c}
\hline\hline
Fit &  RS & M~(MeV) & $\Gamma$/2~(MeV) & $|\gamma_1|$~(GeV) & $|\gamma_2/\gamma_1|$       
\\ \hline\hline
Fit-1A & I & $2356.7\sim 2362.8$ & $0$ & $1.3\sim 6.9$ & $1.03\sim 1.20$   
\\ 
Fit-1A & II & $2316.7\sim 2362.8$ & $0$ & $0.4\sim 10.1$ & $1.14\sim 1.50$  
\\ \hline 
Fit-1B & I & $2357.1\sim 2362.8$ & $0$ & $0.5\sim 6.7$ & $1.05\sim 1.22$  
\\ 
Fit-1B & II & $2316.0\sim 2362.8$ & $0$ & $0.6\sim 10.3$ & $1.12\sim 1.56$  
\\ \hline 
Fit-2A & I & $2345.1_{-41.5}^{+14.7}$ & $0$ &$8.3_{-2.6}^{+2.3}$  & $0.96_{-0.08}^{+0.06}$  
\\ \hline 
Fit-2B & I & $2350.7_{-25.7}^{+9.0}$ & $0$ &$7.7_{-2.0}^{+2.1}$  & $0.83_{-0.06}^{+0.08}$  
\\\hline\hline
\end{tabular}
\end{footnotesize}
\caption{\label{tab.phypoledk}  Poles and their residues obtained at physical meson masses from the $S$-wave coupled-channel $DK$ and $D_s\eta$ scattering amplitudes with $(S,I)=(1,0)$. For Fit-1A and Fit-1B, parts of the parameter configurations within one-sigma uncertainties give bound state poles for the $\ds$ and others give virtual poles. In these cases, we simply show the ranges for the bound- and virtual-state poles obtained for the parameter configurations from Fit-1A and Fit-1B within one-sigma uncertainties. For Fit-2A and Fit-2B, all the parameter configurations within one-sigma uncertainties give bound state poles. The indices of the residues $\gamma_{i=1,2}$ correspond to the channels $DK$ and $D_s\eta$ in order. The physical thresholds of the $DK$ and $D_s\eta$ channels are $2362.8$ and $2516.2$~MeV, respectively. For the definition of  the different RS's, see the text.} 
\end{table}

 \begin{table}[htbp]
 \centering
\begin{footnotesize}
\begin{tabular}{ c  | c c c c}
\hline\hline
Channels&  Fit-1A & Fit-1B &  Fit-2A & Fit-2B
\\ \hline
$a^{(-1,0)}_{D\bar{K}\to D\bar{K}}$ & $-4.53^{\,*}$ & $0.96_{-0.44}^{+1.44}$  & $21.9^{\,*}$ & $0.68_{-0.16}^{+0.17}$ 
\\\hline
$a^{(-1,1)}_{D\bar{K}\to D\bar{K}}$& $-0.18_{-0.01}^{+0.01}$ &  $-0.18_{-0.01}^{+0.01}$& $-0.20_{-0.01}^{+0.01}$ & $-0.19_{-0.02}^{+0.02}$ 
\\\hline
$a^{(2,\frac{1}{2})}_{D_sK\to D_s K}$ & $-0.19_{-0.01}^{+0.01}$&  $-0.19_{-0.01}^{+0.01}$ &$-0.20_{-0.01}^{+0.01}$ & $-0.19_{-0.01}^{+0.01}$ 
\\\hline
$a^{(0,\frac{3}{2})}_{D\pi\to D\pi}$ & $-0.098_{-0.004}^{+0.005}$&$-0.101_{-0.003}^{+0.005}$  &$-0.103_{-0.003}^{+0.003}$ &$-0.099_{-0.004}^{+0.003}$
\\\hline
$a^{(1,1)}_{D_s\pi\to D_s\pi}$&$0.012_{-0.005}^{+0.005}$ & $0.005_{-0.003}^{+0.003}$ &$0.012_{-0.003}^{+0.003}$ & $0.003_{-0.002}^{+0.002}$
\\
$a^{(1,1)}_{DK\to DK}$& $-0.19_{-0.17}^{+0.12}+i\, 0.55_{-0.07}^{+0.08}$&$0.06_{-0.03}^{+0.05}+i\,0.30_{-0.05}^{+0.09}$ &$-0.01_{-0.03}^{+0.05}+i0.39_{-0.04}^{+0.04}$ &$0.05_{-0.03}^{+0.04}+i0.17_{-0.03}^{+0.03}$ 
\\\hline
$a^{(1,0)}_{DK\to DK}$ & $2.16^{\,*}$&$2.36^{\,*}$ &$-1.51_{-2.35}^{+0.72}$ &$-1.87_{-1.98}^{+0.85}$ 
\\
$a^{(1,0)}_{D_s\eta\to D_s\eta}$ &$-0.54_{-0.03}^{+0.06}+i\,0.25_{-0.12}^{+0.17}$&$-0.54_{-0.03}^{+0.07}+i\,0.24_{-0.12}^{+0.15}$ &$-0.39_{-0.03}^{+0.05}+i0.06_{-0.02}^{+0.02}$&$-0.33_{-0.05}^{+0.03}+i0.07_{-0.02}^{+0.02}$
\\\hline
$a^{(0,\frac{1}{2})}_{D\pi\to D\pi}$ &$0.39_{-0.03}^{+0.03}$ &$0.31_{-0.01}^{+0.01}$ &$0.40_{-0.02}^{+0.03}$ & $0.34_{-0.03}^{+0.00}$ 
\\
$a^{(0,\frac{1}{2})}_{D\eta\to D\eta}$ &$-0.50_{-0.06}^{+0.07}+i\,0.27_{-0.15}^{+0.36}$&$0.20_{-0.29}^{+0.10}+i\,0.57_{-0.28}^{+0.62}$ &$0.29_{-0.22}^{+0.15}+i0.61_{-0.26}^{+0.30}$ & $0.16_{-0.06}^{+0.11}+i0.13_{-0.03}^{+0.07}$
\\
$a^{(0,\frac{1}{2})}_{D_s\bar{K}\to D_s\bar{K}}$&$-0.56_{-0.05}^{+0.05}+i\,0.09_{-0.03}^{+0.08}$&$-0.73_{-0.27}^{+0.21}+i\,0.43_{-0.11}^{+0.08}$  & $-0.57_{-0.04}^{+0.06}+i0.35_{-0.07}^{+0.08}$ &$-0.26_{-0.10}^{+0.05}+i0.52_{-0.03}^{+0.06}$
\\
\hline\hline
\end{tabular}
\caption{\label{tab.sl} Predictions of the $S$-wave scattering lengths for various channels obtained at the physical meson masses. The values are given in units of fm. For the entries marked by asterisks, only the central values are given. See the text for details.}
\end{footnotesize}
\end{table}

\section{Conclusions}\label{sect.conclusion} 

In this work we simultaneously analyzed the lattice finite-volume energy levels and the scattering lengths for the scattering of the charmed and light pseudoscalar mesons from Refs.~\cite{Liu:2012zya,Moir:2016srx,Mohler:2013rwa,Lang:2014yfa} within the chiral effective field theory. Several different fit strategies, by using different values for the pion decay constant and including different data sets in the fits, are explored in our study. Through the fits of the lattice data, we fix the values the chiral low-energy constants up to the next-to-leading order and the subtraction constants introduced in the unitarization procedure. The updated values for the low energy constants and subtraction constants are collected in Table~\ref{tab.fitlecs}, which provide a useful starting point for future study on the charmed resonance dynamics in various processes. 

The scattering amplitudes and the resonance poles obtained at the physical meson masses provide the most important outputs of this work. Regarding the resonance poles in the coupled-channel $D\pi, D\eta$ and $D_s\bar{K}$ $S$-wave scattering with $(S,I)=(0,1/2)$, a robust conclusion in our study is that there is pole with the mass around 2100~MeV and the width more than 200~MeV, see Table~\ref{tab.phypoledpi}. Another type of heavier poles lying between 2300~MeV and 2500~MeV, depending on different fits, are also found, with their widths varying from 70~MeV to 200~MeV. According to Fig.~\ref{fig.phasedpiphymass}, the physical $S$-wave $D\pi$ phase shifts and inelasticities with $I=1/2$ show somewhat different behaviors from different fits, specially in the energy region above 2350~MeV. To implement the scattering amplitudes obtained here in the semileptonic charmed meson decays~\cite{Yao:2018tqn} or the phenomenological study of $B$ decays~\cite{Du:2017zvv} may offer us another way to further discriminate the different fits.

For the phase shifts and inelasticities of the $S$-wave coupled-channel $DK$ and $D_s\eta$ scattering with $(S,I)=(1,0)$, we find two different types of solutions. In one solution, i.e. the lower branch of the phase shifts in Fig.~\ref{fig.phasedkphymass}, the $D_{s0}^{*}(2317)$ corresponds to a bound state. While the upper branch of the phase shifts in the former figure implies a virtual state nature of the $D_{s0}^{*}(2317)$. Future lattice simulations with more energy levels in this channel may enable us to discriminate the two different solutions.

\section*{Acknowledgements}
Finite-volume energy levels taken from Ref.~\cite{Moir:2016srx} were provided by the Hadron Spectrum Collaboration --
no endorsement on their part of the analysis presented in the current paper should be assumed. We would like to thank Feng-Kun Guo and De-Liang Yao for valuable discussions. 
This work is funded in part by the NSFC under grants No.~11575052, the Natural Science Foundation of Hebei Province under contract No.~A2015205205.
This work is also partially supported by the Sino-German Collaborative Research Center ``Symmetries and the Emergence of Structure in QCD''
(CRC~110) co-funded by the DFG and the NSFC. JAO would like to thank partial financial support to the MINECO (Spain) and EU grant FPA2016-77313-P.  
UGM and AR acknowledge the support from Volkswagenstiftung under contract No. 93562. 
AR acknowledges the support from Shota Rustaveli National Science Foundation (SRNSF), grant No. DI-2016-26. 
The work of UGM was also supported by the Chinese Academy of Sciences (CAS) President's
International Fellowship Initiative (PIFI) (Grant No. 2018DM0034). LL acknowledges the support from the Key Research Program of the Chinese Academy of Sciences, Grant NO. XCPB09.

\end{document}